\documentclass[%
aip,
apl,
amsmath,amssymb,
preprint,%
floatfix
]{revtex4-1}

\usepackage{graphicx}
\usepackage{dcolumn}
\usepackage{bm}
\usepackage{color}

\usepackage{soul,xcolor}
\setstcolor{red} 

\usepackage{graphicx}
\usepackage{dcolumn}
\usepackage{bm}
\usepackage[mathlines]{lineno}
\usepackage{pdfpages}
\makeatletter
\AtBeginDocument{\let\LS@rot\@undefined}
\makeatother

\begin{document}

\preprint{}

\title[]{Thru-Hole Epitaxy: Is Remote Epitaxy Really Remote?}

\author{Dongsoo Jang}
\thanks{These authors contributed equally: Dongsoo Jang, Chulwoo Ahn, Youngjun Lee}
\affiliation{Department of Physics, Kyung Hee University, Seoul 02447, Korea}

\author{Chulwoo Ahn}
\thanks{These authors contributed equally: Dongsoo Jang, Chulwoo Ahn, Youngjun Lee}
\affiliation{Department of Information Display, Kyung Hee University, Seoul 02447, Korea}

\author{Youngjun Lee}
\thanks{These authors contributed equally: Dongsoo Jang, Chulwoo Ahn, Youngjun Lee}
\affiliation{Department of Physics, Kyung Hee University, Seoul 02447, Korea}

\author{Seungjun Lee}
\affiliation{Department of Physics, Kyung Hee University, Seoul 02447, Korea}

\author{Hyunkyu Lee}
\affiliation{Department of Information Display, Kyung Hee University, Seoul 02447, Korea}

\author{Donghoi Kim}
\affiliation{Department of Information Display, Kyung Hee University, Seoul 02447, Korea}

\author{Young-Kyun Kwon$^{*,}$}
\thanks{email: ykkwon@khu.ac.kr, jaewuchoi@khu.ac.kr, ckim@khu.ac.kr}
\affiliation{Department of Physics, Kyung Hee University, Seoul 02447, Korea}
\affiliation{Department of Information Display, Kyung Hee University, Seoul 02447, Korea}

\author{Jaewu Choi$^{*,}$}
\thanks{email: ykkwon@khu.ac.kr, jaewuchoi@khu.ac.kr, ckim@khu.ac.kr}
\affiliation{Department of Information Display, Kyung Hee University, Seoul 02447, Korea}

\author{Chinkyo Kim$^{*,}$}
\thanks{email: ykkwon@khu.ac.kr, jaewuchoi@khu.ac.kr, ckim@khu.ac.kr}
\affiliation{Department of Physics, Kyung Hee University, Seoul 02447, Korea}
\affiliation{Department of Information Display, Kyung Hee University, Seoul 02447, Korea}



\begin{abstract}
The remote epitaxy was originally proposed to grow a film, which is not in contact but crystallographically aligned with a substrate and easily detachable due to a van der Waals material as a space layer.  Here we show that the claimed remote epitaxy is more likely to be nonremote `thru-hole' epitaxy. On a substrate with thick and symmetrically incompatible van der Waals space layer or even with a three-dimensional amorphous oxide film in-between, we demonstratively grew GaN domains through thru-holes via connectedness-initiated epitaxial lateral overgrowth, not only readily detachable but also crystallographically aligned with a substrate.  Our proposed nonremote thru-hole epitaxy, which is embarrassingly straightforward and undemanding, can provide wider applicability of the benefits known to be only available by the claimed remote epitaxy.
\end{abstract}



\maketitle

\section*{Introduction}

It is a golden rule that epitaxial growth of a crystalline film is allowed only by direct bonding to a crystalline substrate.  Extremely surprising studies reported that a crystalline film was epitaxially grown remotely without direct bonding to an underlying substrate despite an ultrathin defect-free 2D overlayer placed in-between, which was named as remote epitaxy.\cite{Kim_Nature_544_340,Kong_NM_17_999,Jeong_SA_6_eaaz5180}  An earlier study, for instance, claimed that GaN can be remotely grown through only up to two layers of graphene but not even a single layer of $h$-BN on underlying GaN.\cite{Kong_NM_17_999}  This result was explained by the teleported influence of the GaN substrate that was not completely screened.\cite{Kong_NM_17_999} Once the thickness of an inserted 2D material becomes above a critical value, the grown film is no longer crystallographically aligned with the underlying substrate.\cite{Kong_NM_17_999,Jeong_SA_6_eaaz5180}

The remote epitaxy strictly requires not only the defect-free growth of 2D material but also the state-of-the-art transfer with precise layer-number control. Nevertheless, great attention has been paid to remote epitaxy since it has been believed to provide the great benefit of easy separation of the film, which is crystallographically aligned with a substrate.\cite{Kong_NM_17_999,Kim_Nature_544_340,Jeong_SA_6_eaaz5180} Several other studies have also demonstrated the facile detachment of a film taken as evidence of remote epitaxy.\cite{Jeong_Nanoscale_10_22970,Jeong_APL_113_233103,Guo_NL_20_33,Jeong_ACSANM_3_8920,Bae_NM_18_550}  Strictly speaking, however, the easy detachability may not necessarily originate from the `remoteness' of the remote epitaxy, but simply indicates that the binding between the film and space layer/substrate is weaker than the adhesion of the film to a detacher such as a thermal release tape. In fact, the entire remoteness of the remote epitaxy across the interface has never been rigorously verified yet.  The estimated potential fluctuation across the surface of 2D material/substrate given as theoretical evidence of remote epitaxy was only one-tenth of thermal energy at growth temperature, and more importantly, the corresponding potential profile is essentially uncorrelated to that of the bare substrate.\cite{Kong_NM_17_999}  Moreover, it turned out that the overlooked connectedness directly to the substrate through the space layer (Extended Data Fig.~7 of Ref.~[1]) should not have been neglected.  Thus, it is physically more reasonable to presume that the \textit{claimed} remote epitaxy is not likely to be remote.

\begin{figure*}
\includegraphics[width=1.0\columnwidth]{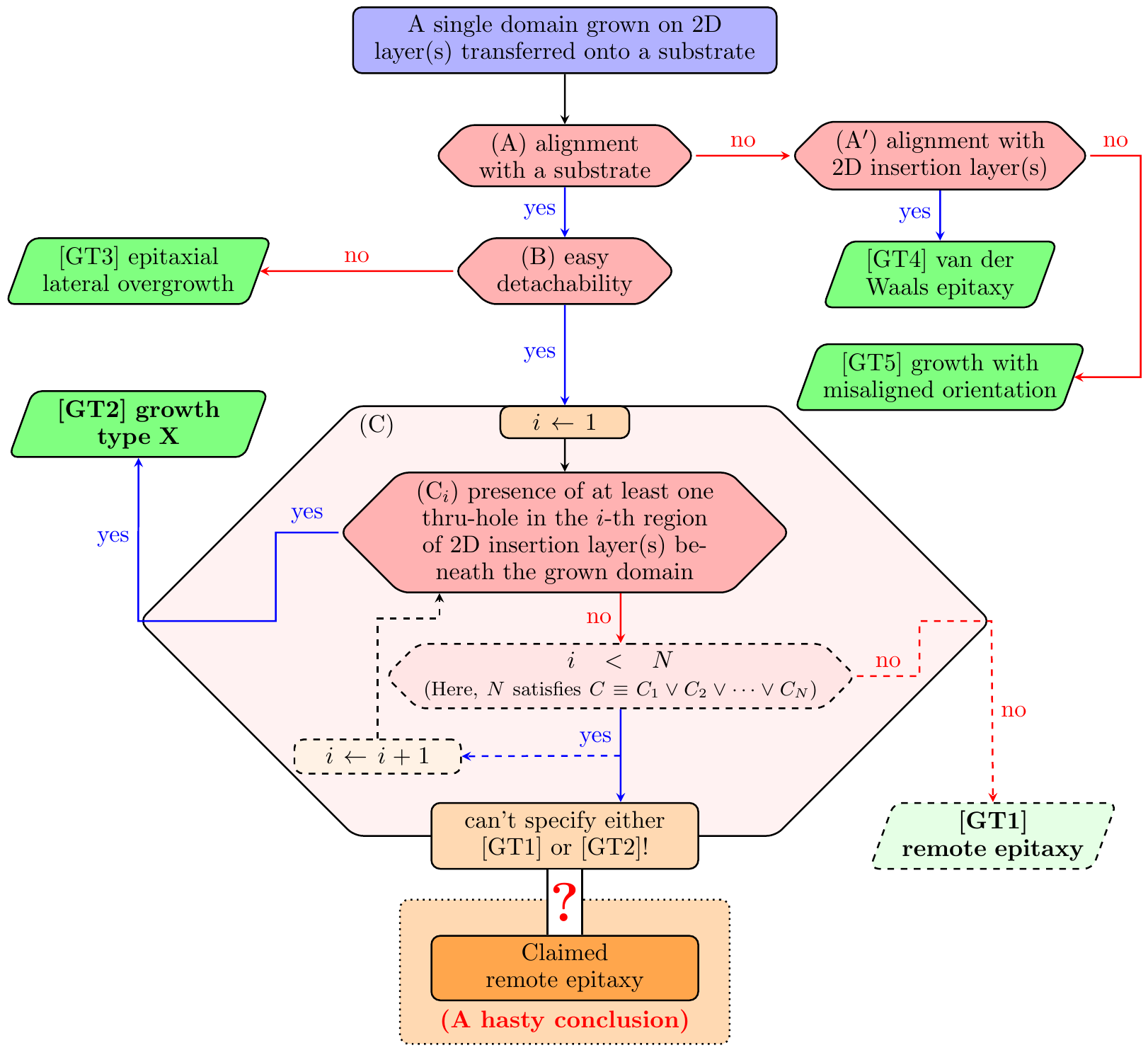}
\caption{\textbf{Logic flow chart of growth types categorization.}  Five different growth types are categorized on basis of alignment, easy detachability, and the existence of thru-holes in a 2D insertion layer beneath the grown domain. Performed processes or flows are denoted by solid boxes or lines, whereas dashed boxes or lines indicate unperformed processes or flows. Note that remote epitaxy was claimed as a growth type in a farfetched manner on basis of insufficient experimental evidence by which [GT1] or [GT2] cannot be specified.}
\label{Logic_flow}
\end{figure*}

In order to support our argument above, we elaborate on the logic related to growth types categorization. Let us consider a single domain grown on a 2D layer transferred onto a single crystalline substrate.  The growth type [GT] for this domain can be categorized into one of the following five as shown in Fig.~\ref{Logic_flow}: [GT1] remote epitaxy, [GT2] growth type X, [GT3] epitaxial lateral overgrowth (ELOG), [GT4] van der Waals epitaxy, and [GT5] growth with misaligned orientation.  These GT's were categorized on basis of alignment (A or A$^\prime$), easy detachability (B), and the existence of at least one thru-hole in the 2D insertion layer beneath the grown domain (C).  The presence of at least one thru-hole in the $i$-th region of the 2D insertion layer beneath the growth domain is labeled as (C$_i$).  Thus, (C) is equivalent to $C_1\vee C_2\vee\cdots\vee C_N$ where $N$ is the number required for the union of 1-st, 2-nd, $\cdots$, $N$-th region to fully cover the \emph{entire} region of 2D insertion layer beneath the growth domain.  On the other hand, the logical negation of (C$_i$) or NOT (C$_i$), denoted as $(\overline{\mathrm{C}_i})$, is the complete absence of thru-hole in the $i$-th region of 2D insertion layer beneath the growth domain.  Likewise, $(\overline{\mathrm{C}})$, indicating NOT (C), is the complete absence of thru-hole in the \emph{entire} region of 2D insertion layer beneath the growth domain.  Of course, these growth types are not necessarily exclusive to one another, and under certain circumstances, mixed growth types may be observed among various domains or within a merged film.  However, for the simplicity of discussion, we assume that one growth type is associated with one single domain.

The necessary and sufficient condition for [GT1] remote epitaxy is the combination of (A) and $(\overline{\mathrm{C}})$.  The verification of $(\overline{\mathrm{C}})$ requires a full series of high-resolution cross-sectional transmission electron microscopy (HR-TEM) images all across the interface, which has never been given yet. This kind of extensive HR-TEM measurements is, however, impractical even for micro-sized domains, so that the verification of $(\overline{\mathrm{C}_1})$, which was typically shown by a \textit{single} HR-TEM image in the previous papers on remote epitaxy, was provided as evidence for remote epitaxy instead of $(\overline{\mathrm{C}})$.  On the other hand, we named any growth type, which satisfies (A), (B), and (C), as [GT2] growth type X, which is definitely not [GT1] remote epitaxy. As can be readily inferred from Fig.~\ref{Logic_flow}, however, a connectedness-initiated epitaxy through very small thru-holes can be a candidate for [GT2] growth type X because HR-TEM in the thru-hole-free region of the domain grown by [GT2] growth type X can reveal that $(\overline{\mathrm{C}_i})$ is satisfied and only a few small thru-holes may allow easy detachability.  In our manuscript, the existence of growth type X was explicitly verified and we identified this growth type X as thru-hole epitaxy. [GT2] Thru-hole epitaxy distinctively refers to the case in which grown domains are not only aligned with a substrate but also easily detached.  Easy detachability is the main distinction of [GT2] thru-hole epitaxy differentiating from conventional [GT3] ELOG.  As can be seen from Fig.~\ref{Logic_flow}, we would like to emphasize that the combination of (A), (B) and $(\overline{\mathrm{C}_1})$ is not enough to specify either [GT1] remote epitaxy or [GT2] growth type X, \textit{i.e.}, thru-hole epitaxy.  In other words, the combination of (A), (B) and $(\overline{\mathrm{C}_1})$ is not sufficient enough to verify the existence of [GT1] remote epitaxy.  Note that remote epitaxy was claimed as a growth type in a farfetched manner on basis of insufficient experimental evidence by which [GT1] or [GT2] cannot be specified.

Furthermore, we show that all the other claimed evidences of remote epitaxy are also evidences of thru-hole epitaxy. (See Supplementary Note1 for more detailed explanation.) Thus, either remote epitaxy or thru-hole epitaxy cannot be specified on basis of all the combined experimental evidences, claimed by remote epitaxy, which are just necessary conditions of remote epitaxy.  Only if $(\overline{\mathrm{C}_1})\wedge (\overline{\mathrm{C}_2})\wedge\cdots\land (\overline{\mathrm{C}_N})$ equivalent to $(\overline{\mathrm{C}})$ were verified, the existence of [GT1] remote epitaxy would be verified.  The flaw of logic in the previous papers on remote epitaxy is that the existence of [GT1] remote epitaxy was claimed to be confirmed only by $(\overline{\mathrm{C}_1})$ instead of $(\overline{\mathrm{C}})$ although other evidences were given. Moreover, our computational simulation revealed that the theoretical evidence of [GT1] remote epitaxy is questionable as described below.  With appropriate justification ((i) logic flow chart, (ii) our experimental results, and (iii) our computational simulation), we claim that the existence of remote epitaxy has not been verified yet and is questionable.

It appears that the original proposers of remote epitaxy recently realized the existence of thru-holes and recognized that the previous experimental evidences for remote epitaxy may not be enough to specify either [GT1] remote epitaxy or [GT2] thru-hole epitaxy.\cite{Kim_ACSNano_15_10587}
In order to find out the role of thru-holes and specify either remote epitaxy or thru-hole epitaxy, they designed and carried out experiments, but they reached a hasty conclusion in which [GT2] was excluded even in the presence of thru-holes.\cite{Kim_ACSNano_15_10587} (See Supplementary Note2 for the more detailed explanation.)

\section*{Results and discussion}

Based on the argument made above, we hypothesized that the claimed remote epitaxy may simply be `thru-hole' epitaxy composed of two processes through nanoscale holes sparsely distributed in the space layer: nucleation on the exposed substrate and lateral growth\cite{Nam_APL_71_2638} over the space layer.  To verify our proposed hypothesis, we demonstratively grew epitaxial GaN domains under various \textit{et mediocri} conditions, which are far from those required by the state-of-the-art growth and transfer indispensable in the claimed remote epitaxy.  By doing so, we reveal that our proposed nonremote thru-hole epitaxy can equally well provide in an unprecedentedly straightforward manner the benefits, which are supposedly accompanied only by the claimed remote epitaxy, such as facile detachability and crystallographic alignment with a substrate. In the following, we first provide computational results showing the unfeasibility of remote epitaxy and then experimental results verifying thru-hole epitaxy.

\subsection*{Computational evidence questioning remote epitaxy\\}

\begin{figure*}
\includegraphics[width=1.0\columnwidth]{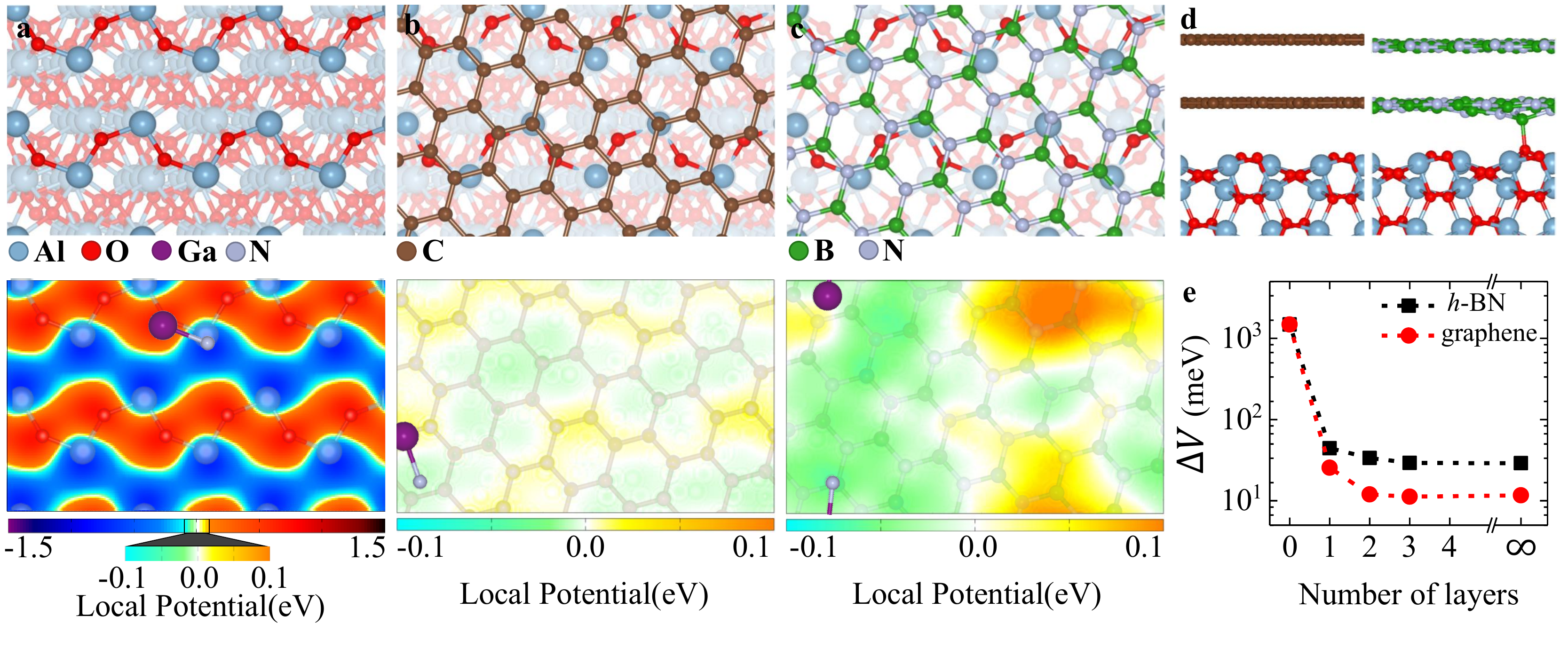}
\caption{\textbf{Structural configurations and surface potential profiles of sapphire with and without graphene or hBN overlayer.} \textbf{a--c} Structural configurations of (\textbf{a}) bare $r$-sapphire, (\textbf{b}) graphene, and (\textbf{c}) $h$-BN monolayer on the $r$-sapphire, and their surface potential profiles evaluated at a distance $d$ from the respective top surfaces. The distance $d$ was chosen to be 2.0~{\AA} and 3.0~{\AA} on the bare $r$-sapphire and each of the overlayers, which are approximately bonding distances between GaN and the respective surfaces. \textbf{d} Side views of bilayers of graphene and h-BN on the $r$-sapphire. \textbf{e} The maximum potential differences $\Delta V$ observed between the Ga and N sites of a Ga-N dimer considered as an initial seed depicted in (\textbf{a--c}).
The infinity indicates the 2D overlayer without the $r$-sapphire substrate. The topmost atomic layer overlaid on each color-coded potential profile is a guide for the eyes. The color bars indicate the potential variation relative to the average potential value set to be zero. Note that the potential variations on the overlayers were color-coded within a much narrower range.  Each 2D space layer/$r$-sapphire is composed of a $2\times3$ rectangular cell of the $r$-sapphire and a rectangular cell modified from a $(\sqrt{13}\times\sqrt{13})R13.9^\circ$ cell of a 2D overlayer, which somewhat mimics a naturally-occurring incommensurate stacking.}
\label{Theory}
\end{figure*}

The remote epitaxy proposed by earlier studies\cite{Kong_NM_17_999,Kim_Nature_544_340,Jeong_SA_6_eaaz5180,Jeong_Nanoscale_10_22970,Jeong_APL_113_233103,Guo_NL_20_33,Jeong_ACSANM_3_8920,Bae_NM_18_550} was validated mainly by the surface potential distribution calculated over a space layer complying with that of a substrate. We discussed some questions raised from the previous calculations in Supplementary Note3. To verify whether such validation is indeed valid, we evaluated the electrostatic potential distributions on various surfaces including not only bare substrates such as sapphire, Si, GaAs, GaN, and LiF but also those substrates with a 2D graphene or $h$-BN overlayer, which will later be denoted as `overlayer/substrate', using the first-principles density functional theory (DFT) calculations. As an exemplary demonstration, we display the surface structures of the bare $r$-sapphire, graphene/$r$-sapphire, and $h$-BN/$r$-sapphire in the top view and their calculated surface potential profiles in Figs.~\ref{Theory}(a--c), respectively. Here we emphasize that a computationally-convenient small commensurate stacking configuration introduced to resolve the lattice mismatch between the substrate and the 2D overlayer instigates an artificially periodic potential fluctuation, which would lead to a misinterpretation, as discussed later. To mimic a naturally-occurring incommensurate stacking, we thus constructed relatively a large supercell structure composed of 2D overlayer/$r$-sapphire.

As shown in the lower part of Fig.~\ref{Theory}a, the bare $r$-sapphire without a space layer generates a huge potential variation, which is attributed to both a strong ionic characteristic of Al-O bonds and an uneven surface configuration of the substrate.  When a monolayer ($n=1$) of the 2D space overlayer, either graphene or $h$-BN, covers the $r$-sapphire, not only is the surface potential variation drastically reduced due to the screening of the space layer, but it does not reflect the shape of the potential profile on the $r$-sapphire substrate, as shown in the lower parts of Figs.~\ref{Theory}b and c.  With one more layer ($n=2$) of the overlayer material, not to mention the complete dissimilarity between its surface potential distribution and that on the bare substrate, the surface potential variation becomes much smaller, being almost close to that over the overlayer itself without the substrate, since the second layer is essentially flat as shown in Fig.~\ref{Theory}d. Furthermore, the potential variations for $n{\geq}3$ cases were found to be almost the same as that of either isolated graphene or $h$-BN, as shown in Supplementary Fig.~1.

To examine whether the remote epitaxy would indeed have been eventuated on defect-free 2D space materials, we estimated the potential difference $\Delta V$ undergone by a Ga-N dimer regarded as a primordial growth seed that should anchor on the 2D overlayer with a similar orientation to on the bare substrate to guarantee the remote epitaxy. Figure.~\ref{Theory}e shows $\Delta V$ as a function of the layer numbers $n$ of 2D graphene and $h$-BN overlayers. Even with $n=1$, $\Delta V$ was calculated to be only $1\sim2$~\% of that on the bare sapphire, and rapidly converged to the value on the 2D overlayer without the substrate when $n{\geq}2$. For comparison, over other substrates, such as Si, GaAs, GaN, and LiF, we evaluated their potential variations at $d=3.0$~\AA, near which the 2D space layer would locate, above their surfaces, to know over which substrate the potential variations emerge through a 2D space layer. As shown in Supplementary Fig.~2, the magnitude of the evaluated potential variation only over the GaN substrate is similar to that over the $r$-sapphire, but those over the other substrates are much smaller.  Thus, the teleportation of substrate potential variations through a 2D space layer is not likely to occur over any kind of substrates as seen in Fig.~\ref{Theory} for $r$-sapphire.

We further investigated the potential variations on the $c$- and $m$-sapphire substrates while increasing the layer number $n$ of the $h$-BN space material from $n=0$ to $n=3$, whose trends are essentially identical to that on the $r$-sapphire case, as shown in Supplementary Fig.~3. To verify the effect of an artificially-generated periodicity caused by the strain applied to forcibly match the lattice mismatch between the substrate and the space layer as mentioned above, we evaluated the surface potential profiles over various stacking configurations of $h$-BN/$c$-sapphire, the combination of which clearly evinces such stacking effect. As displayed in Supplementary Fig.~4, the calculated potential variations are very sensitive to a choice of stacking configurations.  Especially, in small supercell configurations, local potential variations look as if that of the substrate would be reflected over the 2D space layer, but this is a misleading artifact caused by such a forcibly matched stacking to make it commensurate. Therefore, the potential fluctuation of 2D material/substrate does not truly reflect the orientation and periodicity of the underlying substrate if the artifact originating from artificial stacking configurations is simply excluded. For a detailed explanation, see Supplementary Note4.

Based on our calculations, the remoteness of the claimed remote epitaxy\cite{Kong_NM_17_999,Kim_Nature_544_340,Jeong_SA_6_eaaz5180,Jeong_Nanoscale_10_22970,Jeong_APL_113_233103,Guo_NL_20_33,Jeong_ACSANM_3_8920,Bae_NM_18_550} is conceptually and strongly questionable, and thus it should be replaced by a genuine growth mechanism, nonremote thru-hole epitaxy. The proposed thru-hole epitaxy enabled us to grow epitaxial GaN domains, which still exhibit all the benefits from the seemingly remote epitaxy, in the growth regime even prohibited by the claimed remote epitaxy.

\subsection*{Thru-hole epitaxy enabled by connectedness\\}

\begin{figure}
\includegraphics[width=1.0\columnwidth]{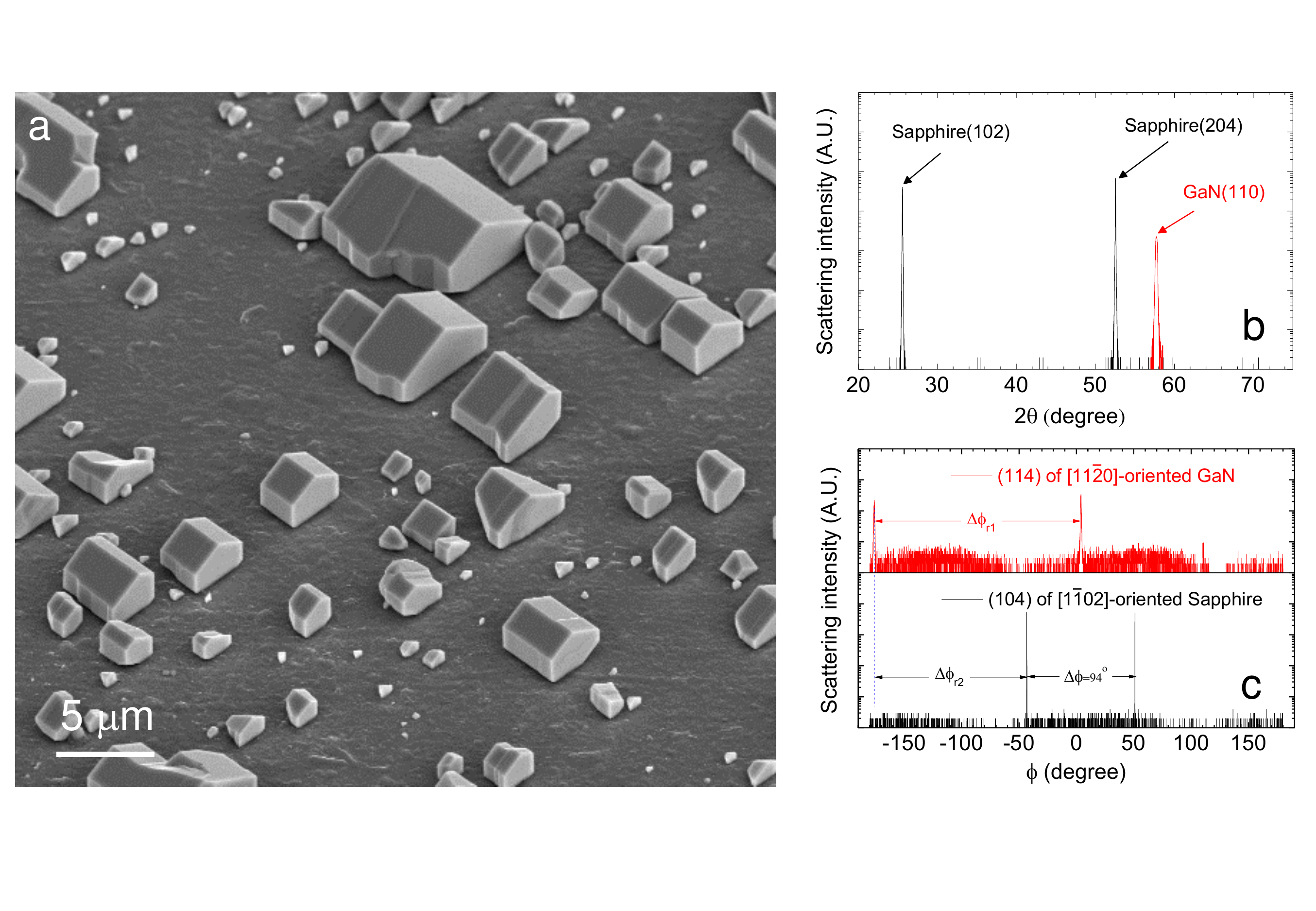}
\caption{\textbf{Crystallographic alignment of GaN domains grown on $h$-BN/$r$-sapphire verified by SEI and XRD.} \textbf{a} An SEI of typical GaN domains grown on $h$-BN/$r$-sapphire. The garble-roof-shaped GaN domains\cite{Shin_JAC_46_443} are aligned in parallel.  \textbf{b--c} X-ray scattering intensity of GaN and $r$-sapphire in (\textbf{b}) $\theta$-2$\theta$ and (\textbf{c}) $\phi$ scan.  The observed  [11\={2}0]-oriented GaN domains are exactly aligned with $r$-sapphire as if they were grown on the bare $r$-sapphire substrate as indicated by $\Delta\phi_{r1}$=180$^{\rm \circ}$ and $\Delta\phi_{r2}$=133$^{\rm \circ}$.\cite{Chen_JJAP_42_L818,Imer_JCG_306_330}
}
\label{GaN_on_differently_oriented_sapphire}
\end{figure}

To verify our proposed growth mechanism, we grew GaN domains on a thick and polycrystalline $h$-BN space layer transferred onto $r$-, $m$- and $c$-oriented sapphire substrates.  The thickness of $h$-BN was deliberately chosen by multiple transfers to make the influence of a substrate completely negligible and to entirely exclude the remoteness of the claimed remote epitaxy.  It was surprisingly observed that those GaN domains grown on a $h$-BN/$r$-sapphire substrate exhibit the [11\={2}0]-orientation with a garble-roof-shape as if directly grown on the bare $r$-sapphire substrate, as shown in a secondary electron image (SEI) of  Fig.~\ref{GaN_on_differently_oriented_sapphire}a.  Even more surprisingly, those [11\={2}0]-oriented GaN domains were all in-plane aligned with one another on the thick and randomly-oriented polycrystalline  $h$-BN layer, (Supplementary Fig.~5) clearly indicating that the growth result is not what is expected in van der Waals epitaxy but exactly what would be expected in the claimed remote epitaxy.  The crystallographic alignment of those GaN domains with the $r$-sapphire substrate is confirmed by X-ray scattering measurement as shown in Figs.~\ref{GaN_on_differently_oriented_sapphire}b and c.  Likewise, similar results were observed on $h$-BN/$c$- and $m$-sapphire. (Supplementary Fig.~6) We observed the exact one-to-one correspondence in Bragg peaks of GaN domains respectively grown on $h$-BN/sapphire and on bare sapphire. (Supplementary Fig.~7) It seemed that we dramatically extended the applicable growth regime of the claimed remote epitaxy, resulting from the crystallographic information of sapphire teleported to GaN even across a thick $h$-BN space layer.

\begin{figure*}
\includegraphics[width=0.85\columnwidth]{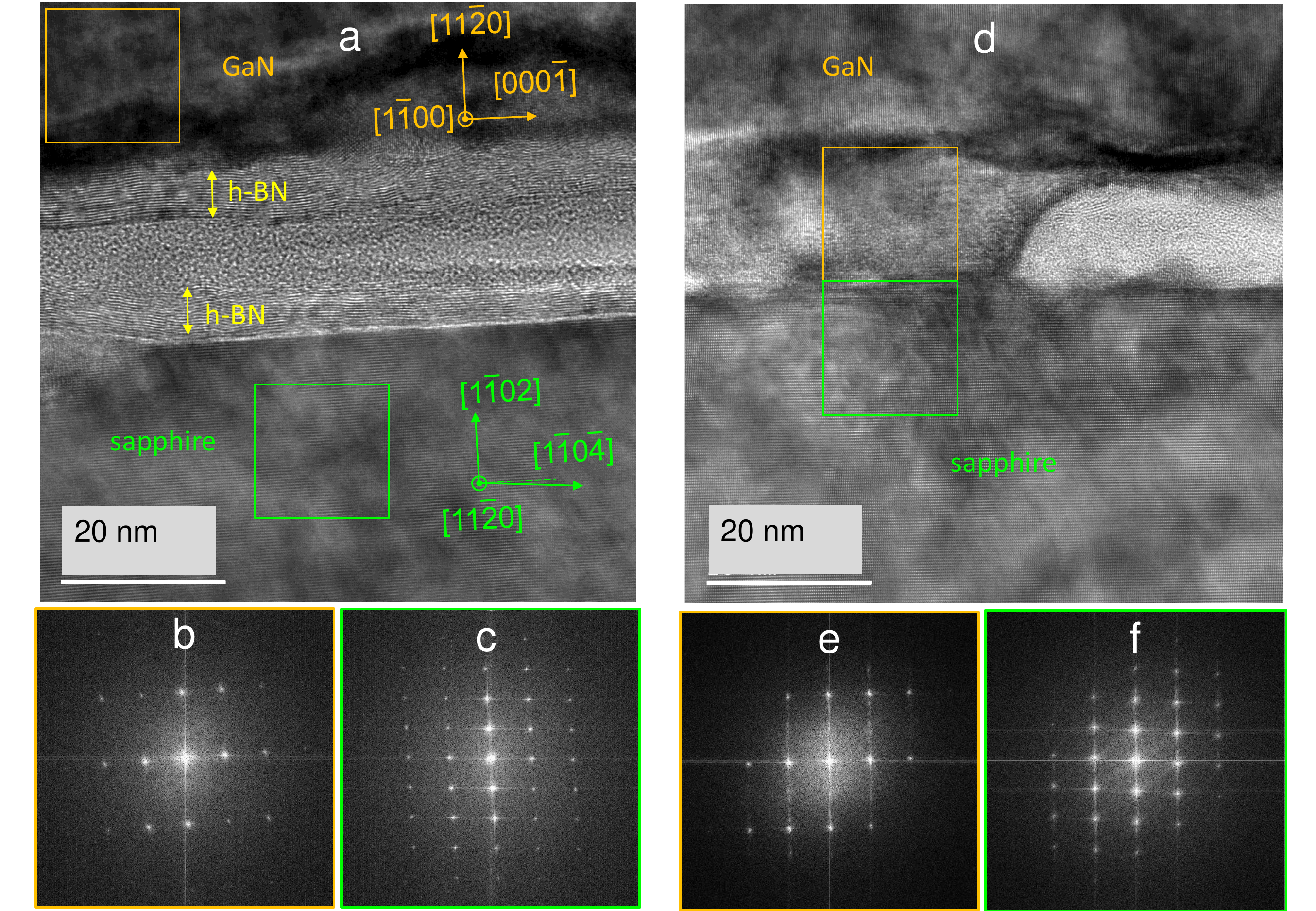}
\caption{\textbf{TEM and FFT verifying the connectedness.} \textbf{a} High-resolution cross-sectional TEM image, taken from the region enclosed by a dashed cyan box shown in Supplementary Fig.~8a and b.  \textbf{b--c} The fast Fourier transforms (FFTs) of [11\={2}0]-oriented GaN and $r$-sapphire enclosed respectively by the (\textbf{b}) orange and (\textbf{c}) green boxes marked in (\textbf{a}).  \textbf{d} High-resolution cross-sectional TEM image, taken from the region enclosed by a dashed yellow box shown in Supplementary Fig.~8b. \textbf{e--f} GaN is seen to be grown directly on $r$-sapphire indicating the connectedness achieved, (Supplementary Fig.~9a) and their crystallographic alignment was verified by FFTs in (\textbf{e}) and (\textbf{f}).  Note that there are crystalline $h$-BN layers, marked in (\textbf{a}), which embed carbon-based amorphous materials. (Supplementary Fig.~9b) They are thought to originate from the residue of PMMA used for $h$-BN transfer and do not make any influence on the thru-hole epitaxy.  The slight tilt of a line along the laterally aligned Bragg peaks of [11\={2}0]-oriented GaN with respect to that of $r$-sapphire shown in FFT is due to the characteristics of crystallographic orientation of $r$-sapphire, not due to the stress relaxation.
}
\label{TEM}
\end{figure*}

Then, what caused this astonishing crystallographic alignment of GaN domains with underlying sapphire substrates even covered by the thick and polycrystalline $h$-BN?  To answer this question, we extensively performed TEM measurement.  Figure~\ref{TEM}a shows a high-resolution cross-sectional TEM image revealing the interfacial region of GaN and $r$-sapphire with a thick $h$-BN space layer in-between. (Supplementary Fig. 8) Enigmatically, the fast Fourier transformation (FFT) analysis shown in Figs.~\ref{TEM}b and c revealed that GaN, which is [11\={2}0]-oriented, right above thick $h$-BN was aligned with $r$-sapphire consistent with XRD measurements shown in Figs.~\ref{GaN_on_differently_oriented_sapphire}b and c.  It was unveiled that the origin of this striking crystallographic alignment of GaN is thru-hole epitaxy.  As shown in Fig.~\ref{TEM}d, [11\={2}0]-oriented GaN was directly connected to the $r$-sapphire substrate through thru-holes in $h$-BN in spite of multiple transfers of $h$-BN. (Supplementary Fig. 9) As shown in Fig.~\ref{TEM}e and f, FFTs near thru-holes through which GaN and sapphire are connected show the crystallographic alignment of GaN with underlying sapphire.  It can be naturally inferred that GaN nucleated on exposed sapphire through thru-holes must have laterally grown over a thick $h$-BN space layer as seen in Fig.~\ref{TEM}a.  Our experimental results, absolutely unexpected by the claimed remote epitaxy, were made possible not because the remoteness was mysteriously enhanced but because the connectedness was securely established.

It is the first direct evidence of our proposed thru-hole epitaxy, which is applicable to any systems and also even to the claimed remote epitaxy.  It is farfetched that the claimed remote epitaxy was attributed to remoteness, but it is rather reasonable to connectedness secured by thru-hole epitaxy, as verified by our computational results described above.  In a previous study, as a matter of fact, this same kind of connectedness was also observed on small areas but unfortunately disregarded as a  minor effect.\cite{Kim_Nature_544_340}  Being above the critical layer thickness or losing the connectedness is the reason why the claimed remote epitaxy fails.  Thus, we claim that the connectedness serving as a control parameter is utilized as a true criterion not only for the claimed remote epitaxy but also for the thru-hole epitaxy.  We showed how to control the extent of the connectedness by intentionally adjusting the thickness of and deliberately not improving the quality of a 2D space layer. (Supplementary Fig.~10)  We emphasize that the better the quality of a 2D space layer the lower the density of thru-holes.  As a result, the weaker the connectedness the smaller the critical value of the thickness, above which the claimed remote epitaxy or the thru-hole epitaxy is forbidden.\cite{Kong_NM_17_999,Jeong_Nanoscale_10_22970} Therefore, the thru-hole epitaxy of crystallographically aligned GaN on $h$-BN/sapphire does not require the state-of-the-art perfection of $h$-BN at all.

The readily facile detachability of a grown film crystallographically aligned with an underlying substrate is the exclusive benefit of the claimed remote epitaxy.  To check whether the grown film by thru-hole epitaxy is readily detachable as well, we carried out the detachability experiment. Supplementary Fig.~11 shows that GaN domains were readily detached simply by using a thermal release tape although they were connected to the substrate by thru-holes. (Supplementary Fig.~12)  We emphasize that there is a critical connectedness below which all the benefits of the claimed remote epitaxy can be obtainable.  Unlike the stringent requirement imposed by the claimed remote epitaxy, the connectedness below such critical connectedness is readily achievable. (Supplementary Fig.~13)

\subsection*{Extended demonstration of thru-hole epitaxy\\}

\begin{figure*}
\includegraphics[width=1.0\columnwidth]{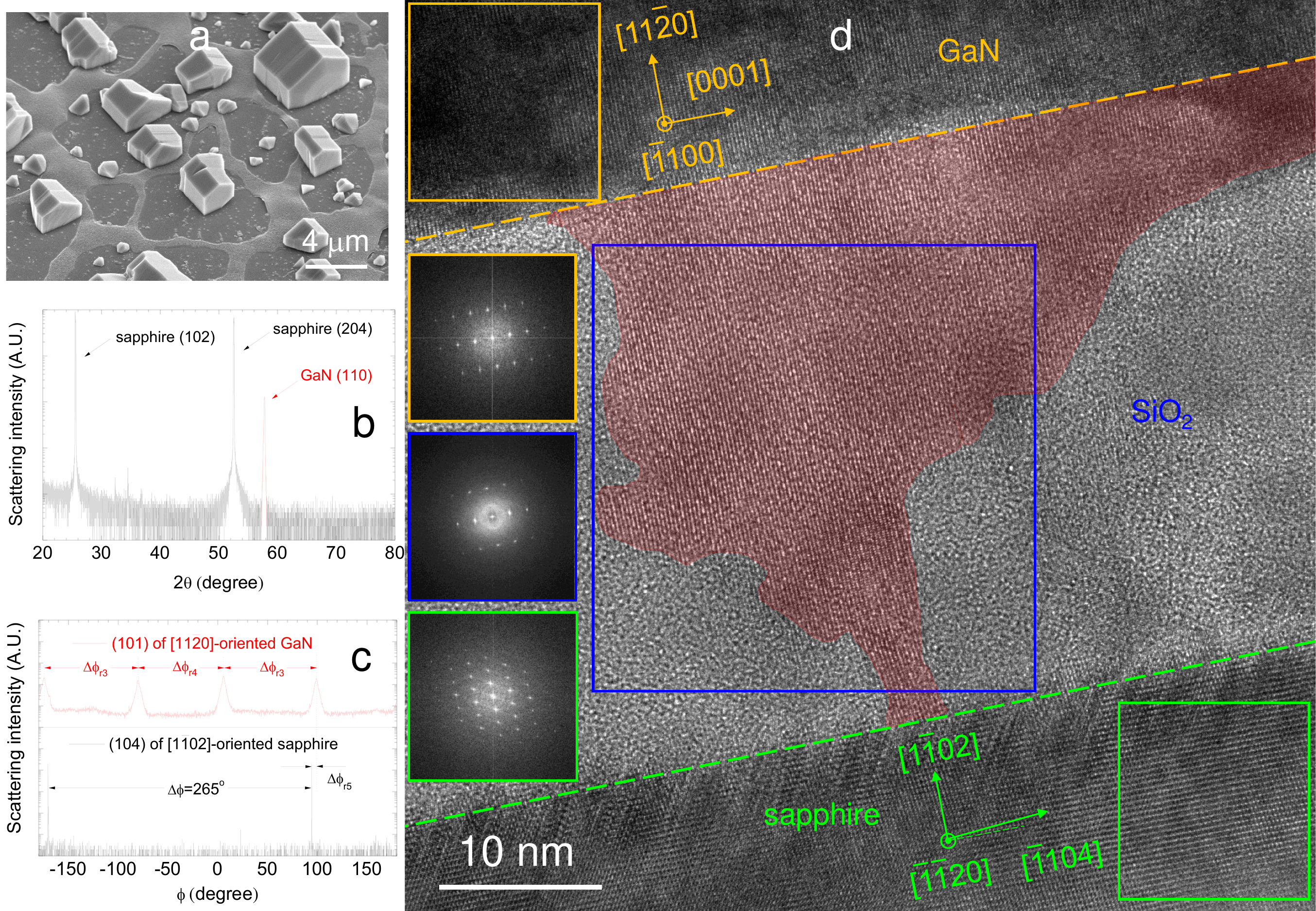}
\caption{\textbf{Compelling evidences of thru-hole epitaxy through a SiO$_2$ space layer.} \textbf{a} SEI of in-parallel aligned GaN domains with garble-roof shape grown on $h$-BN/SiO$_2$/$r$-sapphire.  \textbf{b--c} X-ray scattering intensity of GaN and $r$-sapphire in (\textbf{b}) $\theta$-2$\theta$ and (\textbf{c}) $\phi$ scan.  Those [11\={2}0]-oriented GaN domains are still crystallographically aligned with $r$-sapphire in spite of 50-nm-thick SiO$_2$ space layer verified by $\Delta\phi_{r3}$=94$^{\rm \circ}$, $\Delta\phi_{r4}$=86$^{\rm \circ}$, and $\Delta\phi_{r5}$=5$^{\rm \circ}$.  \textbf{d} High-resolution cross-sectional TEM image, taken from the region enclosed by a dashed red box in Supplementary Fig.~14a and b, and FFTs for various regions.  TEM and FFTs of the regions enclosed by the orange, blue, and green boxes confirm the connectedness by thru-hole epitaxy (shaded in red) and crystallographic alignment.
}
\label{TRT_transfer}
\end{figure*}

Our experimental results showed that the connectedness by thru-holes is crucial to achieving the crystallographic alignment of a film with a 2D-layer-covered substrate.  As long as the connectedness is maintained, the thru-hole epitaxy was found to be valid regardless of either the type or the thickness of a space layer.  Our proposed mechanism might, however, be challenged by the speculation that the growth was partly initiated over the $h$-BN regions the thickness of which was less than a critical value of the claimed remote epitaxy.  To unambiguously exclude the speculation, we introduced an additional 50-nm-thick SiO$_2$ space layer between $h$-BN and $r$-sapphire to form SiO$_2$/$r$-sapphire with and without $h$-BN overlayer, which should completely suppress the remoteness.  Subsequently, nanoscale openings were produced in SiO$_2$ to achieve connectedness.  This experimental configuration fully guarantees the thru-hole epitaxy but completely eliminates the possibility of the remote epitaxy.  As expected from the thru-hole epitaxy, GaN domains grown in these experimental configurations were crystallographically aligned with $r$-sapphire as shown in Figs.~\ref{TRT_transfer}(a--c). This result indicates that we successfully achieved the connectedness by intentionally creating nanoscale openings, (See Methods section) verified by TEM analysis as shown in Fig.~\ref{TRT_transfer}d and Supplementary Fig.~14.  The thru-hole epitaxy is robustly manifested even with a SiO$_2$ space layer.

Moreover, we also successfully detached those [11\={2}0]-oriented GaN domains from $h$-BN/SiO$_2$/$r$-sapphire simply by using a thermal release tape as shown in Supplementary Fig.~15.  Here we would like to emphasize that the crystallographic alignment is associated with the connectedness, whereas $h$-BN or any 2D van der Waals space layer plays a role in not transferring the crystallographic information through but only in allowing the film grown above to be readily detachable.

We have shown that ostensibly `remote' epitaxy based on remoteness is evinced by the nonremote thru-hole epitaxy originating from connectedness. Our proposed thru-hole epitaxy mechanism maintains every advantage (e.g., undemanding detachability, crystallographical alignment with an underlying substrate, independence of space layer symmetry) of the remote epitaxy in an embarrassingly straightforward and undemanding manner. It was also demonstrated that our method can be readily extended to hBN/substrate even with a thick SiO$_2$ film in-between without compromising the advantages described above. This growth behavior can open possibilities for the great advantage of detachable heteroepitaxial film growth with no constraint on the state-of-the-art transfer perfection of 2D materials and the thickness of which should be less than a few layers imposed by the claimed remote epitaxy.


\newpage

\section*{Methods}

\begin{description}

\item[Processes of the thru-hole epitaxy.]
The thru-hole epitaxy consists of the following processes, growth and transfer of $h$-BN, and growth of GaN, which were carried out with no special optimization.  Despite no optimization, we were able to accomplish the thru-hole epitaxy to the extent demanded, indicating that state-of-the-art perfection is not necessary for the thru-hole epitaxy.

  \item[Growth of $h$-BN.] A polycrystalline $h$-BN thin film was grown on a Cu foil by chemical vapor deposition (CVD) at 1000$^{\rm \circ}$C with hydrogen (50 sccm) and argon (100 sccm) as a carrier gas as shown in Supplementary Fig.~5. Growth was carried out with ammonia borane (NH$_3$BH$_3$) as a precursor for two hours after two-hour annealing of the Cu foil.  The thickness of $h$-BN was roughly estimated to be 2--4~nanometer by transmission electron microscopy.  The chemical identification of $h$-BN is shown in Supplementary Fig.~9.

  \item[Transfer of $h$-BN.] This CVD-grown $h$-BN was transferred onto $r$-, $c$-, and $m$-sapphire substrates by applying a wet transfer method with poly (methyl methacrylate) (PMMA). PMMA was spin-coated over $h$-BN grown on a Cu foil, which was then etched with FeCl$_3$. PMMA/$h$-BN was transferred onto the desired substrate, and then that was rinsed twice with deionized water.  After that, the PMMA was removed by acetone, and $h$-BN/substrate was cleaned by isopropyl alcohol.  The extent of the connectedness was intentionally controlled by changing the number of transfers made.  For multiple transfers of $h$-BN, the transfer steps listed above were simply repeated.

  \item[Creation of thru-holes in SiO$_2$.]
  We intentionally created the connectedness by introducing thru-holes in a SiO$_2$ space layer.  Thru-holes were produced in SiO$_2$ during the pre-heating step of the GaN growth by utilizing FeCl$_3$, which is known to thermally decompose SiO$_2$.\cite{Jang_CAP_16_93}

    \item[Growth of GaN.] GaN was grown on $h$-BN/sapphire or $h$-BN/SiO$_2$/sapphire by using hydride vapor phase epitaxy at 960$^{\rm \circ}$C with HCl (10 sccm) flown over metal Ga and NH$_3$ (1500 sccm) carried by N$_2$.

  \item[Electron microscopy, focused ion beam, and EDS.] Transmission electron microscopy (TEM) and selected area diffraction (SAD) measurements were performed by using a Thermo Fisher Titan\textsuperscript{TM}80-300 microscope operated at 300~kV.  The available point resolution is better than 1~{\AA} at the operating acceleration voltage. TEM images were recorded by using a charge-coupled camera (Gatan, Oneview095).  Scanning TEM (STEM) \& energy dispersive spectroscopy (EDS) analysis with Super-X EDS was made by using a Thermo Fisher Talos F200X at 200~kV with a probe size of $\sim$1~nm.  Secondary electron images (SEIs) were taken by using Hitachi S-4700.  For the preparation of TEM samples, focused ion beam (Hitachi, NX5000) was utilized.

  \item[Computational simulation.] First-principles calculations were performed using the density functional theory\cite{Kohn1965} as implemented in Vienna \textit{ab initio} simulation package (VASP)\cite{Kresse1996}. The electronic wavefunctions were expanded by plane wave basis with a kinetic energy cutoff of 520~eV. We employed the projector-augmented wave pseudopotentials\cite{Blochl1994,Kresse1999} to describe the valence electrons, and treated exchange-correlation functional within the generalized gradient approximation of Perdew-Burke-Ernzerhof.\cite{Perdew1996}  Interlayer interactions were incorporated with Grimme-D2 van der Waals correction.\cite{grimme-d2} To reduce long-range interactions from neighboring cells located along the out-of-plane or growth direction, we included a sufficiently large vacuum region of 20~\AA. The Brillouin zone (BZ) of each structure was sampled using a separation of 0.04~\AA$^{-1}$ $k$-point mesh according to the Monkhost-Pack scheme.\cite{MP} The structures were relaxed until all forces became smaller than 0.02~eV/{\AA}. The exchange-correlation potential was excluded in potential fluctuation maps because exchange-correlation potential on a long distance from the surface is incorrect within standard DFT.\cite{Kong_NM_17_999,LVTOT}  Various 2D overlayer/sapphire supercell structures were constructed to avoid artificially generated periodic potentials.

\end{description}

\section*{Acknowledgments}
We gratefully acknowledge financial support from the Korean government (MSIT, MOE) through the National Research Foundation (NRF) of Korea (NRF-2019R1A2C1005417, NRF-2019R1F1A1063643, NRF-2020R1F1A1050725, NRF-2020R1A5A6017701,\\ NRF-2021R1A5A1032996, BK21 FOUR Program). Some portion of our computational work was done using the resources of the KISTI Supercomputing Center (KSC-2020-CRE-0011).

\section*{Author Contributions}
Y.L. and S.L. carried out the computational simulation, supervised by Y.-K.K.  C.A. performed the growth and transfer of $h$-BN, supervised by J.C.  D.J carried out the growth and transfer of GaN with assistance from H.L. and D.K., supervised by C.K.  Data analysis and figure preparation of XRD and TEM were performed by D.J. and C.K.  The manuscript was written by Y.-K.K., J.C., and C.K. with assistance from all the authors.

\section*{Competing interests}
The authors declare no competing interests.

\section*{Data availability}
All data are available in the main text or the supplementary information.

\newpage



\section*{Supplementary material}

**Supplementary Information for**

**Thru-Hole Epitaxy: Is Remote Epitaxy Really Remote?**

Dongsoo Jang,[1, a)] Chulwoo Ahn,[2, a)] Youngjun Lee,[1, a)] Seungjun Lee,[1] Hyunkyu Lee,[2] Donghoi Kim,[2] Young-Kyun Kwon*,[1, 2, b)] Jaewu Choi*,[2, b)] and Chinkyo Kim*,[1, 2, b)]

[1)] *Department of Physics, Kyung Hee University, Seoul 02447, Korea*

[2)] *Department of Information Display, Kyung Hee University, Seoul 02447, Korea*

---

[a)] These authors contributed equally: Dongsoo Jang, Chulwoo Ahn, Youngjun Lee

[b)] email: ykkwon@khu.ac.kr, jaewuchoi@khu.ac.kr, ckim@khu.ac.kr

## Supplementary Note1.  EVIDENCES OR FEATURES CLAIMED BY CLAIMED REMOTE EPITAXY FROM THE VIEWPOINT OF THRU-HOLE EPITAXY

**Absence of thru-hole in a limited region of an interface:**

Due to the very nature of HR-TEM, if a sample for cross-sectional HR-TEM is made in the region free from thru-holes, a single image of HR-TEM will show no sign of direct bonding between a grown film and an underlying substrate. This can easily be observed if a thru-hole-connected region is sparsely populated across the interface. In other words, it is hard to spot thru-holes-connected regions through HR-TEM investigations because they are located several micrometers far away from one another with a very small size ranging from several to a few dozen nanometers. So, the absence of thru-hole provided only by the *partial* investigations using HR-TEM cannot be regarded as sufficient proof for the [GT1] remote epitaxy. However, the outcome of claimed remote epitaxy can be easily and self-consistently explained by [GT2] thru-hole epitaxy without borrowing or resorting to any concept of [GT1] remote epitaxy. (See Fig. 1 in the main manuscript for the growth types classification such as [GT1] and [GT2].)

**Crystallographic alignment of a grown film with an underlying substrate:**

The crystallographic alignment is a consequence of epitaxial growth involving direct bonding between a film and an underlying substrate. So, this evidence can be readily understood in terms of [GT2] thru-hole epitaxy as well and is only a necessary condition for [GT1] remote epitaxy.

**Crystallographic alignment limited by layer number of 2D insertion material:**

The crystallographic alignment claimed in claimed remote epitaxy was observed only on monolayer or bilayer graphene. Such alignment in this limited condition was regarded as the main feature of claimed remote epitaxy. This can be also easily and self-consistently explained by thru-hole epitaxy as follows. Even a state-of-art 2D layer transferred onto a target substrate has some unavoidable holes, which can serve as nucleation spots for thru-

hole epitaxy, although the size of holes may vary from monovacancy to a-few-micrometers. The stacking of 2D layers containing holes would decrease the number density of thru-holes, so that the number density of potential nucleation spots would decrease. One important factor here is that how fast the number density of thru-holes decreases with an increasing number of stacking. If the quality of the 2D layer is excellent, only a few stacking would immediately block all the holes whereas thru-holes still survive even after stacking several times if it is mediocre. In any case, if the number density and size of thru-holes get smaller than critical values, [GT2] thru-hole epitaxy becomes less dominant over [GT5] growth with misaligned orientation.

**Ionicity dependence of crystallographic alignment:**

In claimed remote epitaxy, ionicity was regarded as a key factor to obtain crystallographic alignment. Such aligned domains of ionic material can be readily explained by thru-hole epitaxy as long as the size and number density of thru-holes are larger than their critical values for crystallographic alignment. These critical values vary with materials properties such as ionicity. For example, if two different materials with and without ionicity are separately grown on the same 2D insertion layer on the substrate made of their respective materials, adatoms of the nonionic material would be less attracted toward thru-holes by the exposed substrate than those of the ionic material. That is because the range of attractive interaction from the ***exposed area*** of the nonionic material is much shorter than that of the ionic material. Thus, the formation of misaligned domains on the 2D insertion layer would be more probable with the nonionic material than the ionic material because the actual size and number density of thru-holes are smaller than critical values for the nonionic material but larger than those for the ionic material. In other words, nonionic material would be more likely to be grown as [GT5] growth with misaligned orientation than ionic material on the same 2D insertion layer. It should be noted that this kind of ionicity-dependence of interaction range is not typically observed in conventional ELOG because the size of an individual hole or opening area is so sufficiently large that stochastic diffusion of adatoms toward opening areas is more dominant.

**Easy detachability:**

It should be noted that easy detachability does not necessarily suggest a complete absence of direct bonding between a film and an underlying substrate. Instead, it simply indicates that the adhesive force between the grown film and a thermal release tape is large enough to break the binding force between the grown film and a space layer/substrate. It can be easily inferred that increasing the layer number of 2D insertion material would decrease the number density of thru-holes, so that detachability would be improved as well. Moreover, the detachability would be enhanced by reducing the size of thru-holes. In both situations, however, [GT5] growth with misaligned orientation begins to coexist with [GT2] thru-hole epitaxy. On the other hand, as the size and number density of thru-holes become(s) larger the detachment of a grown film or even a single domain by a thermal release tape would become less possible and eventually impossible with the crystallographic alignment enhanced, indicating that [GT3] ELOG becomes dominant.

# Supplementary Note2.  COMPETITION BETWEEN TWO DIFFERENT GROWTH TYPES

It came to our attention that the original proposers of [GT1] remote epitaxy seemingly also recognized recently that their experimental evidences may not be enough to specify either [GT1] remote epitaxy or [GT2] thru-hole epitaxy.[1] By assuming the existence of [GT1] remote epitaxy as default even though it has not been proved yet, they tried to investigate whether [GT1] remote epitaxy would have been a much more dominant growth mechanism than [GT2] thru-hole epitaxy. They came to rule out [GT2] thru-hole epitaxy, which is the opposite conclusion to ours. In their experimental configuration, they chose a nonionic material, Ge, to exclude the effect of [GT1] remote epitaxy. Then, they assumed that the existence of pinholes would result in [GT2] thru-hole epitaxy, which was supposed to be a dominant growth type consequently resulting in a single crystalline film without considering the possibility of [GT5] growth with misaligned orientation at all. However, the grown film was polycrystalline, (Fig. 4(b,d,f) of in the paper[1] quoted above) so they excluded [GT2] thru-hole epitaxy as a dominant growth mechanism over [GT1] remote epitaxy. In their experiment and interpretation, there were two inappropriate assumptions. The first inappropriate assumption was that the existence of pinholes would allow [GT2] thru-hole epitaxy to be dominant over [GT5] growth with misaligned orientation for a nonionic material. (We already explained why the first assumption was inappropriate in the paragraph starting with 'Ionicity dependence of crystallographic alignment' in Supplementary Note1.) The second inappropriate assumption was that the less dominance of [GT2] thru-hole epitaxy over [GT5] growth with misaligned orientation for a nonionic material can be generally applied to the case in which [GT2] thru-hole epitaxy and [GT1] remote epitaxy could be found at the same time. This second assumption was of course inappropriate because the dominance of one growth type over another cannot be determined on basis of the dominance of one growth type over the other. Thus, they should not have concluded from their observation of misaligned domains of nonionic material (Fig. 4(b,d,f) of in the paper[1] quoted above) that [GT5] was more dominant than [GT2] in any case. The less dominance of [GT2] over [GT5] is simply due to the nonionic character of a material. In contrary to the first inappropriate assumption, it should be noted that [GT5] can coexist with or even be dominant over [GT2] or [GT4] among multiple domains or a merged film although there are pinholes on the

2D insertion layer. In order to exclude the possibility of remote epitaxy, they should have used a thicker 2D insertion layer instead of nonionic material. If they had tried with either ionic or partially ionic material with thicker 2D insertion layer to exclude the possibility of [GT1], they would have been able to observe the aligned domains with a substrate, *i.e.,* the dominance of [GT2] thru-hole epitaxy over [GT5] growth with misaligned orientation. That is exactly what we have observed and reported in our manuscript. They recognized and raised an important issue of thru-hole epitaxy, which has been disregarded in comparison with remote epitaxy, but they designed an inappropriate experiment to check their hypothesis (thru-hole epitaxy is less dominant than remote epitaxy) and then misinterpreted their results. As a consequence, they incorrectly concluded that remote epitaxy is more dominant over thru-hole epitaxy.

## Supplementary Note3.   MISINTERPRETATION IN THE PREVIOUS COMPUTATIONAL RESULTS FOR REMOTE EPITAXY

There are several papers reporting van der Waals (vdW) epitaxy on $h$-BN or graphene/graphite. Nevertheless, it is very challenging to achieve such vdW epitaxy or to grow 3D single-crystalline films on dangling-bond-free 2D materials,[2] because of small potential fluctuation and the absence of nucleation centers. To overcome this limitation in most cases, various defects playing roles as nucleation centers were introduced on a 2D surface using surface modification techniques.[3,4] Moreover, vdW epitaxy has been observed relatively more often on $h$-BN than on graphene because the potential fluctuation on the latter surface is even smaller than that on the former surface. Another important point is whether the potential profile reflects the symmetry of underlying substrate or 2D material for remote or vdW epitaxy, respectively, or not. As clearly shown in our manuscript, the teleported potential fluctuation does not reflect the symmetry of the underlying substrate, whereas the surface potential profile of $h$-BN or graphene does show its complete symmetry. We also agree with the reviewer that the formation of a large nucleate with a radius larger than the critical radius is an important key issue for epitaxy, and our estimated potential fluctuation across the surface of 2D material/substrate is even larger than that of vdW epitaxy. However, we would like to point out two points: (i) the structural configurations of large nucleates formed during the growth process would be quasi-random and thus their symmetry would not be necessarily consistent with the symmetry of substrate; and (ii) even if there is a large nucleate that has the same symmetry of substrate by accident, it is unlikely that such nucleate settles on 2D material with crystallographic alignment to the underlying substrate, because the potential profile of 2D material/substrate does not reflect the symmetry of the underlying substrate as mentioned above and in our manuscript.

Some of the previous papers on remote epitaxy showed that their potential/charge profiles reflect the characteristics of the underlying substrates. It turned out that such results had been obtained only under certain constrained conditions as described in the following. In the paper by Kong *et al.*[5], which is one of the original papers explaining the concept of remote epitaxy, they did not show the *total* potential distribution $U_{\text{tot}}$ of 2D material/substrate, which is the true potential governing growth processes. Instead, they intentionally calculated a potential distribution defined as $U = U_{\text{tot}} - U_{\text{2D}}$, to reflect mainly substrate contribution

by subtracting the 2D material contribution, $U_{2D}$. We would like to emphasize again that the true potential governing growth processes is not this potential difference, but the total potential. Moreover, they used small supercell sizes to enforce commensurability between 2D material and substrate by introducing large strain, resulting in the creation of an artificial symmetry reflecting that of the underlying substrate. Such structural constraint is, of course, unavoidable since a periodic system is required in calculations. That is why we considered various stacking configurations (relatively different orientations and supercell sizes) of 2D material/substrate to exclude constraint-induced artifacts in our calculation. As shown in Fig. 2 in our main manuscript and Supplementary Fig. 4, $U_{tot}$ depends strongly on the stacking configuration (relative orientation and supercell size) and thus does not follow the symmetry of the underlying substrate. Another work[6] showed the charge density with a small supercell instead of potential distribution. In that paper, they computed charge density using $\rho = \rho_{tot} - \rho_{2D}$ to extract the substrate contribution, rather than the total charge density $\rho_{tot}$ of 2D material/substrate, which is a true charge density responsible for the growth process. Therefore, the existence of vdW epitaxy cannot be direct evidence for remote epitaxy, the concept of which is still not regarded to be validated by DFT calculations, but rather strongly questionable.

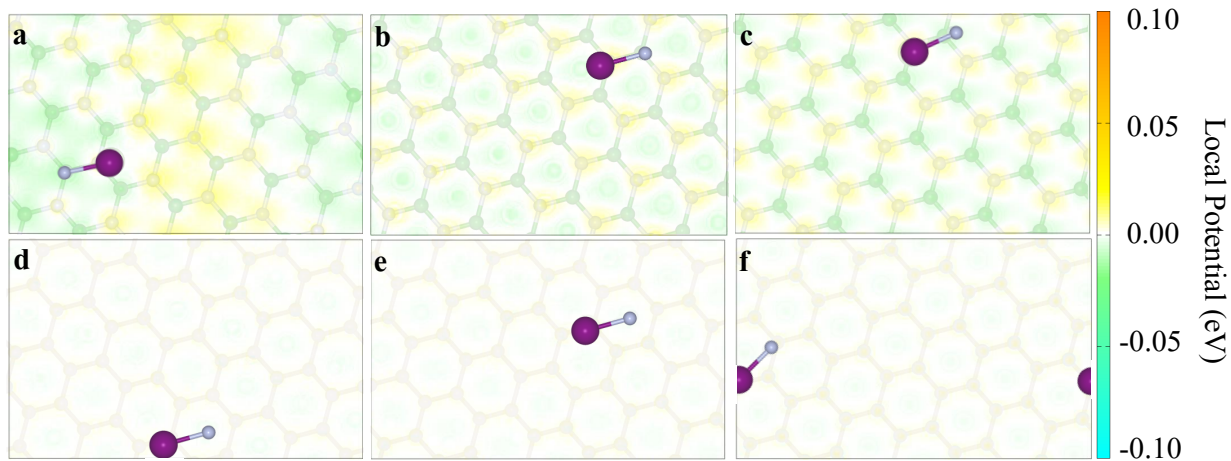

Supplementary Fig. 1: **Surface potential profiles over graphene and $h$-BN multilayers on $r$-sapphire.** **a–f** Surface potential profiles evaluated at $d = 3.0$ Å from the respective top surfaces for (**a**) bi-, (**b**) tri-, and (**c**) hex-layers of $h$-BN, and (**d**) bi-, (**e**) tri-, and (**f**) hex-layers of graphene, all of which are on $r$-sapphire. $\Delta V$ shown in Fig. 2**e** in the main text was evaluated where the Ga-N dimer is placed. The topmost atomic layer overlaid on each color-coded potential profile is a guide for the eyes. The color bar shows the potential variation relative to the average potential value set to be zero. Note that the indistinguishable color variation in (**d–f**) indicates that the potential variations on graphene overlayers are negligibly small.

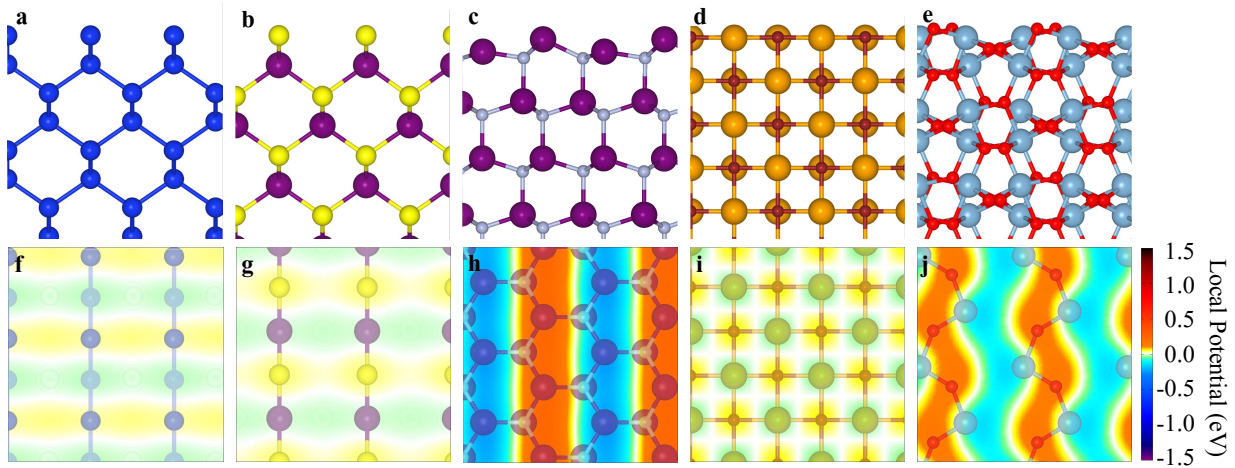

Supplementary Fig. 2: **Surface potential profiles over various substrates. a–e** Side views of (**a**) Si, (**b**) GaAs, (**c**) GaN, (**d**) LiF, and (**e**) $r$-sapphire. **f**–**j** The color-coded surface potential variations calculated at 3.0 Å above the corresponding top surface. The topmost atomic layer overlaid on each color-coded potential profile is guide for the eyes. The color bar shows the potential variation relative to the average potential value set to be zero. Blue, purple, yellow, grey, ocher, brown, red, and skyblue spheres indicate Si, Ga, As, N, Li, F, O, and Al atoms, respectively.

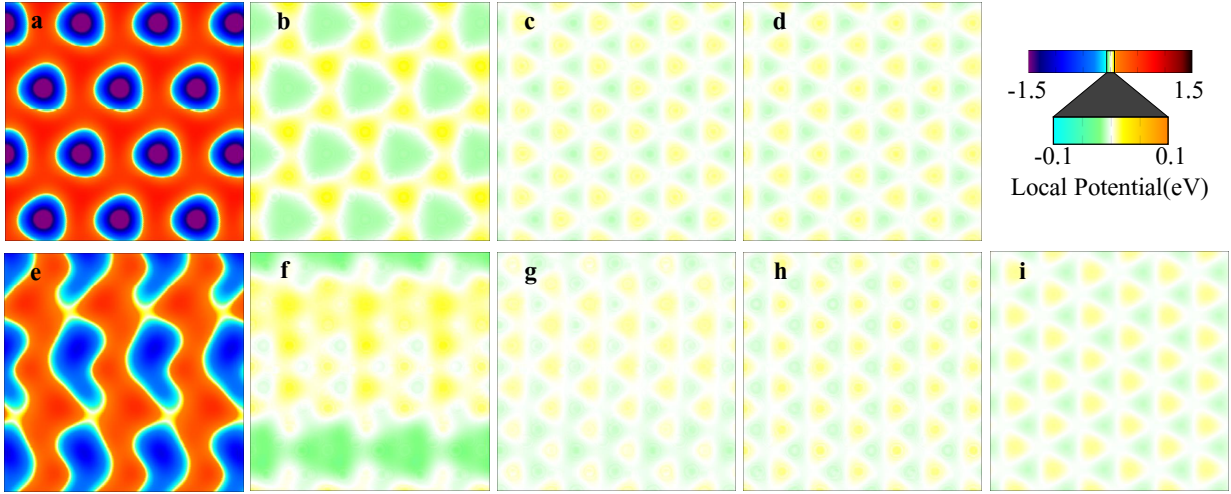

Supplementary Fig. 3: **Surface potential profiles of $h$-BN on $c$- and $m$-sapphire substrate. a–i** Surface potential profiles evaluated at a distance $d$ from the respective top surfaces for (**a**) bare $c$-sapphire, (**b**) mono-, (**c**) bi-, and (**d**) tri-layer $h$-BN on $c$-sapphire, and (**e**) bare $m$-sapphire, (**f**) mono-, (**g**) bi-, (**h**) tri-, and (**i**) hex-layer $h$-BN on $m$-sapphire, respectively. The distance $d$ was chosen to be 2.0 Å on the bare $c$- and $m$-sapphire and 3.0 Å on each of the overlayers, which are approximately bonding distances between GaN and the respective surfaces. When the number of layers of $h$-BN is larger than one, the calculated potential profiles on both $c$- and $m$-sapphire are almost the same as that on the 6 layers of $h$-BN implying a negligibly small contribution of sapphire substrates. The color bars indicate the potential variation relative to the average potential value set to be zero. The potential variations on the $h$-BN overlayers were color-coded within a much narrower range. Note that the symmetry of potential variation, observed in (**b**), similar to that of bare $c$-sapphire is not actual but fictitious due to a lattice distortion caused by the choice of a supercell as described in Supplementary Fig. 4.

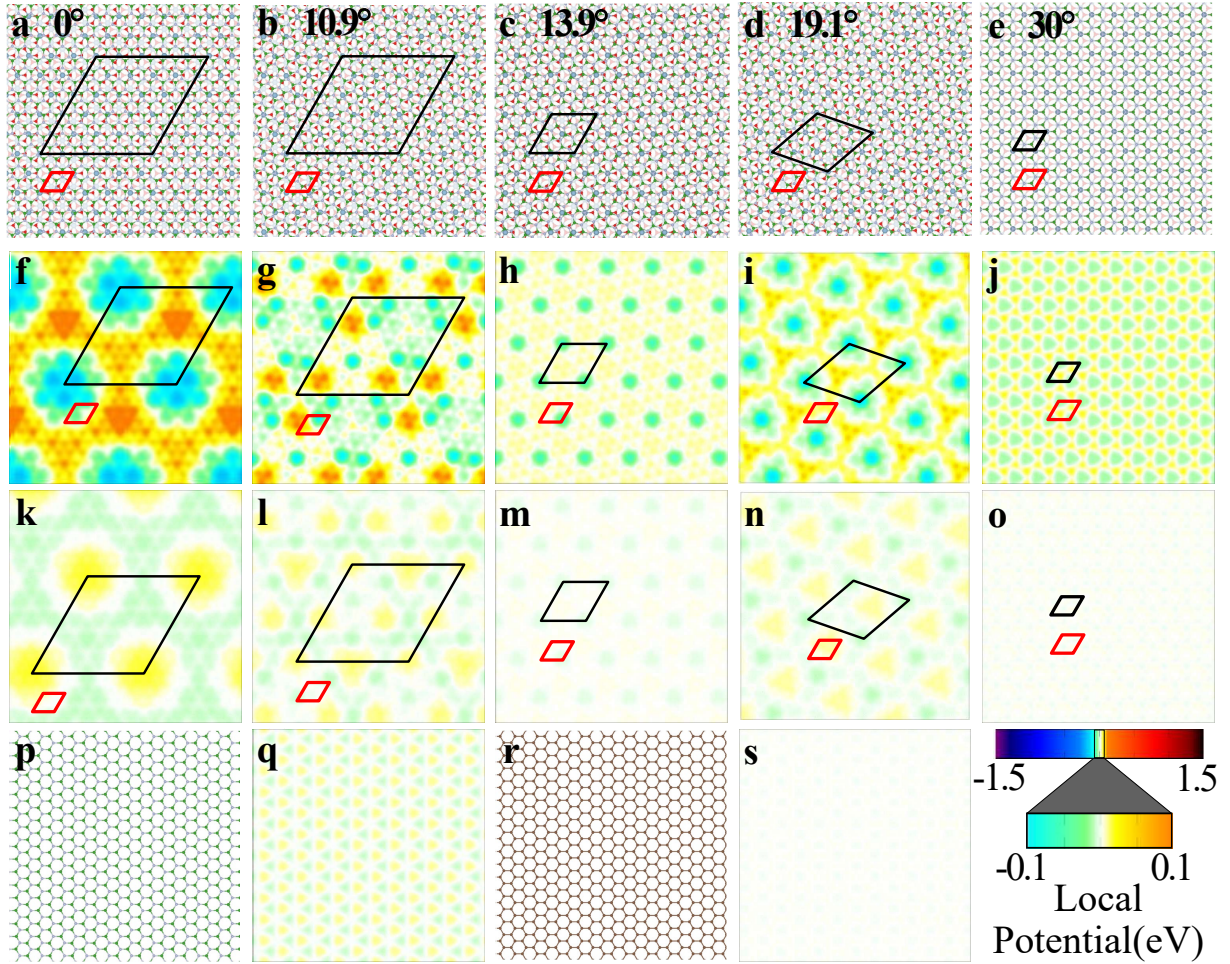

Supplementary Fig. 4: **Misleading artifacts caused by the choice of a supercell and the stacking configuration.** (**a**–**e**) Top views of relaxed structures for various configurations of $h$-BN/$c$-sapphire heterostructure. The numbers in degree indicate the angle between the $h$-BN zigzag direction and the $[10\bar{1}0]$ crystallographic direction of $c$-sapphire. (**f**–**j**) Surface potential profiles calculated at $d = 3.0$ Å above the top surface of the structures shown in (a–e). (**k**–**o**) Surface potential profiles calculated over the graphene, instead of $h$-BN, with the same configurations shown in (a–e). **p**–**s** For comparison, over an isolated $h$-BN monolayer shown in (**p**), we also calculated its surface potential profile as shown in (**q**), as well as for an isolated graphene monolayer as shown in (**r**) and (**s**). The color bars indicate the potential variation relative to the average potential value set to be zero. The supercell of the combined 2D insertion layer and substrate is represented as a black parallelogram while a primitive unit cell of the underlying $c$-sapphire substrate is represented as a red parallelogram. It is clear that the surface potential profiles depend strongly on the stacking configuration as well as the supercell size, resulting in misleading artifacts, especially in small supercell configurations.

12

**Supplementary Note4.   ARTIFACTS CAUSED BY THE CHOICE OF SUPERCELLS**

The potential orientations do not appear to change with the relative rotation in Supplementary Fig. 4, but in fact it does in case of 19.1° rotation as shown in Supplementary Fig. 4 (i). In addition, as can be clearly seen, the side of a black parallelogram has a different length from that of a red parallelogram for each case of different stacking configurations, indicating that the periodicity changes with the relative rotation. Even in these cases, the supercells were created by enforcing artificial commensurability between $h$-BN and $c$-sapphire. For more realistic and practical situations, there would be no artificial commensurability, so that the change of potential orientation and periodicity with the relative rotation between $h$-BN and r-sapphire would be more dramatic. The fact that the potential of the combined system of 2D material/substrate does not follow the orientation and periodicity of the underlying substrate becomes more vivid in cases where 2D material has a symmetry different from that of the substrate, for example, $h$-BN on $r$-sapphire or $m$-sapphire as shown in Fig. 2 in the main manuscript and in Supplementary Fig. 1 and Supplementary Fig. 2. Therefore, the potential fluctuation of 2D material/substrate does not truly reflect the orientation and periodicity of the underlying substrate if the artifact originating from artificial stacking configurations is simply excluded.

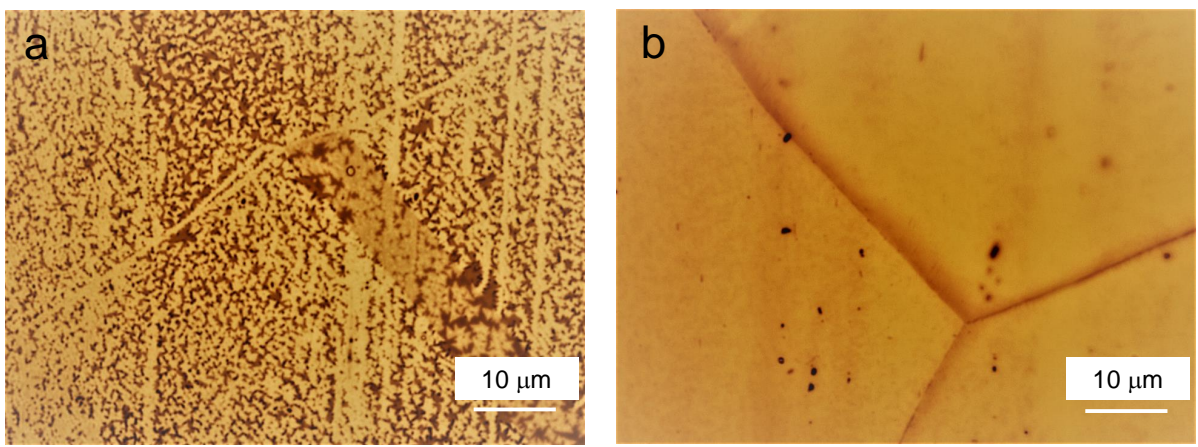

Supplementary Fig. 5: **Polycrystalline $h$-BN grown on a Cu foil.** **a–b** Optical microscopy image of (**a**) 10-min- and (**b**) 2-hr-grown $h$-BN on a polycrystalline Cu foil after 1-min-oxidation at 200°C for a clear distinction between h-BN covered area and exposed Cu foil. The 10-min-grown $h$-BN domains with a typical triangular shape were misaligned with one another, implying that $h$-BN was polycrystalline. On the other hand, the 2-hr-grown h-BN fully covered a Cu foil as shown in (**b**). Note that the color correction was made for clarity.

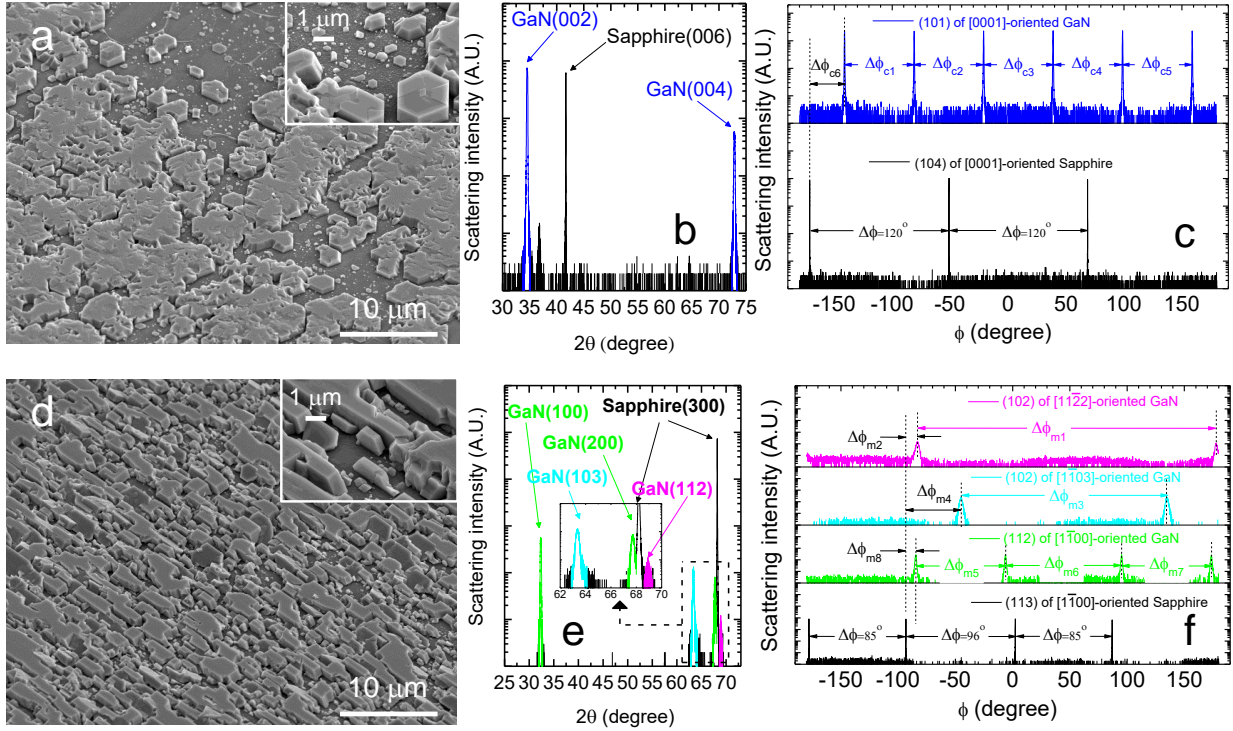

Supplementary Fig. 6: **Crystallographically aligned GaN domains grown on $h$-BN/$c$-sapphire and on $h$-BN/$m$-sapphire.** **a–f** (**a**) Secondary electron image (SEI) and (**b, c**) XRD data of GaN domains grown on $h$-BN/$c$-sapphire as well as (**d–f**) on $h$-BN/$m$-sapphire. It is clear that the crystallographic alignment of GaN domains was determined by the underlying sapphire substrates. XRD $\theta$-$2\theta$- and $\phi$-scan data reveal that the orientation of GaN grown on $h$-BN/$c$-sapphire is [0002] whereas that on $h$-BN/$m$-sapphire is [1$\bar{1}$00], [1$\bar{1}$03], and [11$\bar{2}$2]. The observed orientation is exactly the same as the preferred orientation of GaN grown on bare $c$- and $m$-oriented sapphire substrates[7–15] veified by $\Delta\phi_{c1}=\Delta\phi_{c2}=\Delta\phi_{c3}=\Delta\phi_{c4}=\Delta\phi_{c5}=60°$, $\Delta\phi_{c6}=30°$, $\Delta\phi_{m1}=261°$, $\Delta\phi_{m2}=10°$, $\Delta\phi_{m3}=180°$, $\Delta\phi_{m4}=48°$, $\Delta\phi_{m5}=79°$, $\Delta\phi_{m6}=101°$, $\Delta\phi_{m7}=79°$, and $\Delta\phi_{m8}=8°$. There is a (101) Bragg peak of GaN near $2\theta = 37°$ the intensity of which is several orders smaller than that of the $c$-GaN (002) Bragg peak.

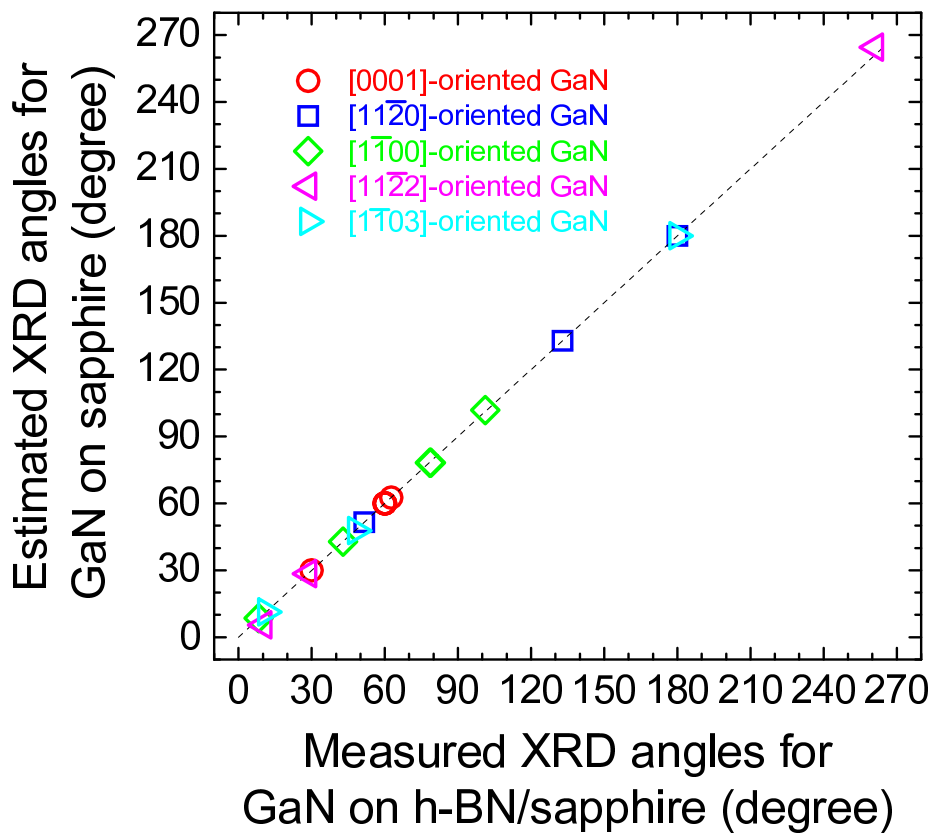

Supplementary Fig. 7: **Consistency between crystallographic alignments of GaN domains grown on $h$-BN/sapphire and bare sapphire.** Experimentally measured XRD angles for GaN domains grown over $h$-BN on $r$-, $c$-, and $m$-sapphire substrates shown in Fig. 3 in the main text and Supplementary Fig. 6 vs. the estimated XRD angles for GaN grown on their corresponding bare sapphire substrates. This consistency verifies the crystallographic alignments of GaN with the underlying sapphire substrates in spite of a $h$-BN space layer.

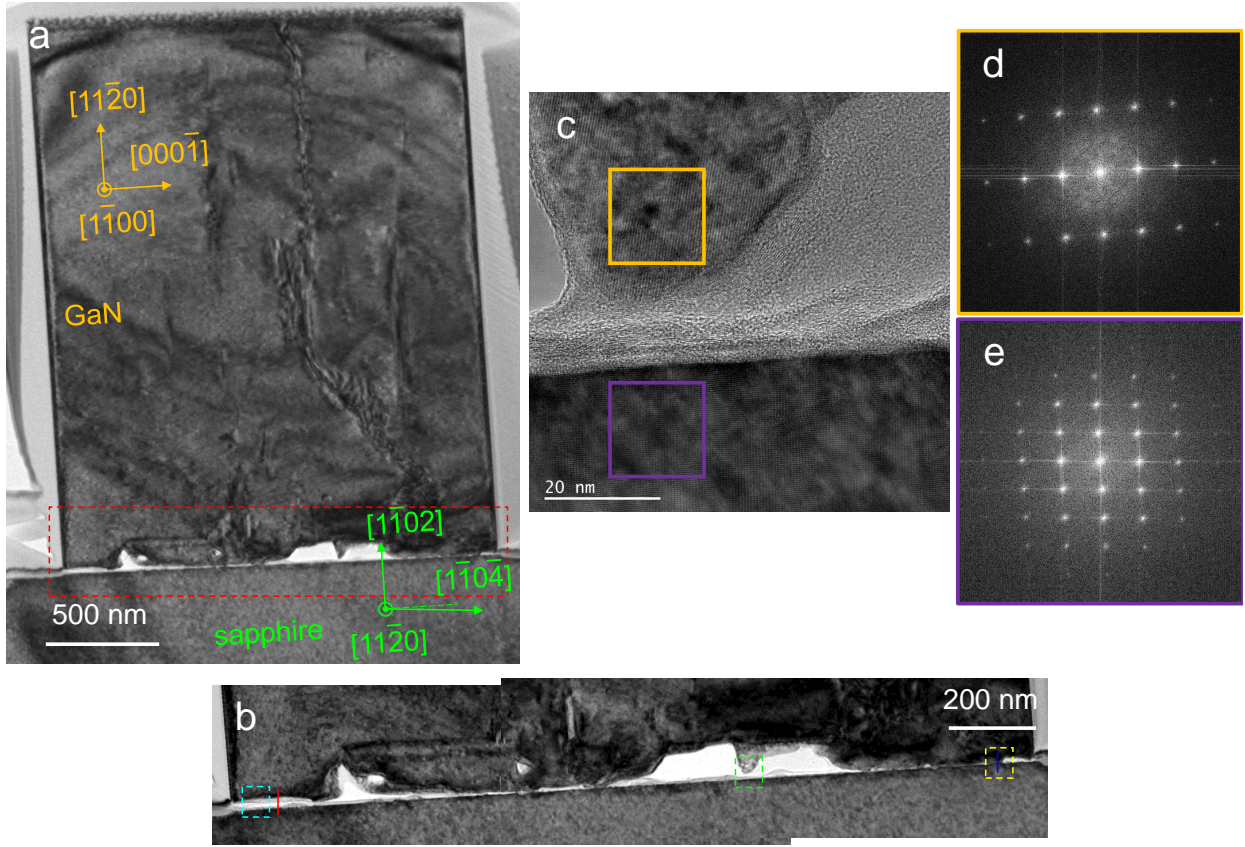

Supplementary Fig. 8: **Direct evidence of thru-hole epitaxy. a** Cross-sectional TEM image of a single domain of GaN grown on $h$-BN/$r$-sapphire. **b** Magnified stitched images of the interface region enclosed by the red dashed box shown in (**a**). **c** Magnified TEM image of the region enclosed by the green dashed box shown in (**b**). The high-resolution TEM image, taken from the regions enclosed by the cyan and yellow dashed boxes, is shown in Figs. 4(**a**) and (**d**) in the main text. The fast Fourier transforms (FFTs) of (**d**) the GaN and (**e**) the $r$-sapphire regions enclosed respectively by the orange and violet boxes in (**c**). The FFTs reveal that [11$\bar{2}$0]-oriented GaN is aligned with $r$-sapphire, suggesting the lateral overgrowth initiated by the thru-hole epitaxy.

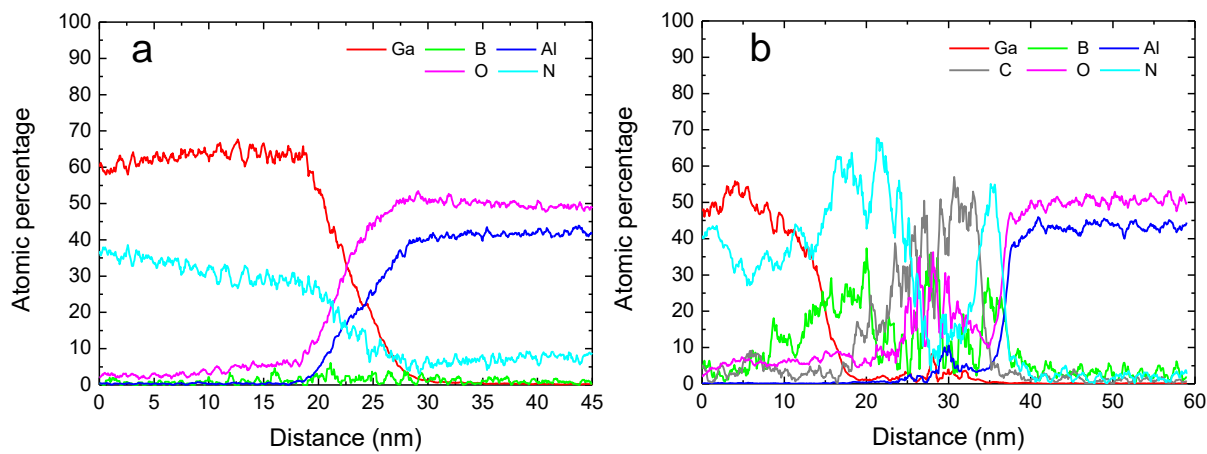

Supplementary Fig. 9: **Chemical analysis across the interface. a** Energy-dispersive X-ray spectroscopy (EDS) line scan taken along the blue line within the yellow dashed box where the connectedness was established as shown in Supplementary Fig. 8b and Fig. 4d in the main text. There is no sign of $h$-BN across the interface indicating the connectedness. **b** EDS line scan, taken along the red line near the cyan dashed box where GaN and sapphire were separated by space layer material as shown in Supplementary Fig. 8b and Fig. 4a in the main text. The existence of $h$-BN across the interface was chemically confirmed. The amorphous material between the $h$-BN layers shown in Fig. 4a in the main text turns out to be carbon-based.

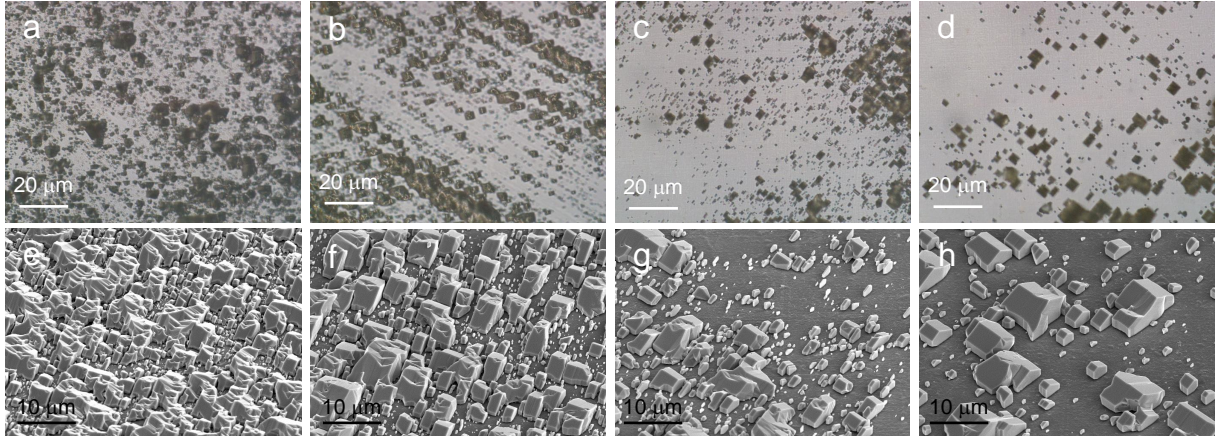

Supplementary Fig. 10: **Control of the extent of the connectedness by adjusting the thickness of a $h$-BN space layer.** **a**–**d** Optical microscopy images of and **e**–**h** SEIs of $[11\bar{2}0]$-oriented GaN domains grown over $h$-BN transferred onto $r$-sapphire substrates (**a**, **e**) once, (**b**, **f**) four, (**c**, **g**) six, and (**d**, **h**) eight times to adjust the thickness of a $h$-BN space layer. It is clear that the nucleation density of GaN domains aligned in parallel with one another decreased with the reduced connectedness resulting from the increased number of transfers of $h$-BN.

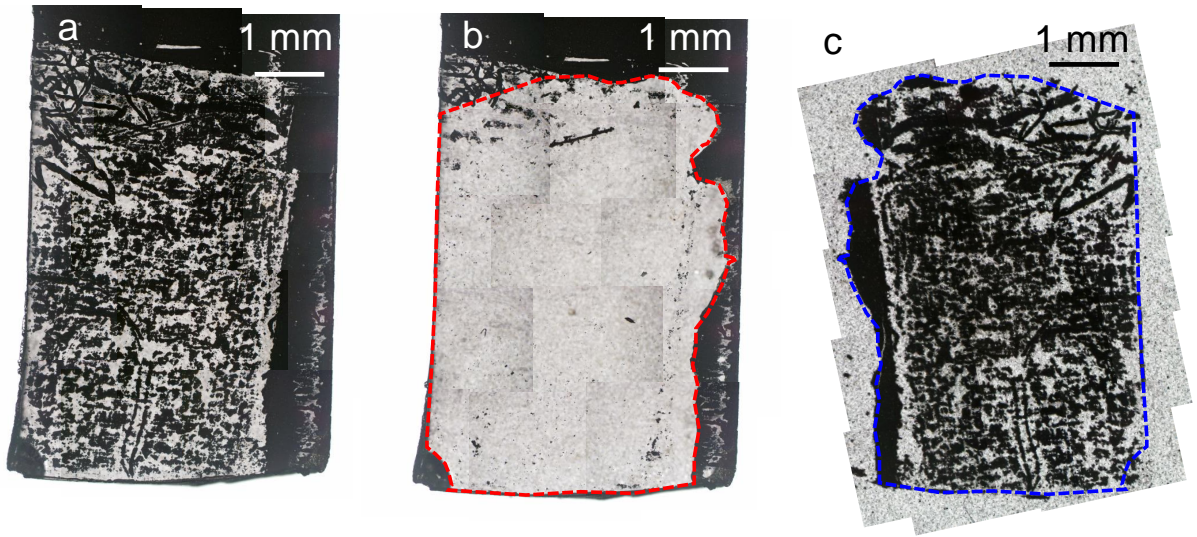

Supplementary Fig. 11: **Facile detachment of GaN domains.** **a**–**c** Stitched optical microscopy images of (**a**) as-grown GaN domains on six-time transferred $h$-BN space layer on an $r$-sapphire substrate, (**b**) the $r$-sapphire substrate after GaN domains were detached by using a handy thermal release tape, and (**c**) the detached GaN domains on the thermal release tape. Although GaN is transparent, those GaN domains shown in the normally back-illuminated optical microscopy images in (**a**–**c**) look unusually dark due to their garble-roof shape as described in Supplementary Fig. 12. The regions enclosed by the red and blue dashed lines represent the exposed $r$-sapphire and detached GaN. Note that there are dark-looking GaN domains with a garble-roof shape outside the region enclosed by the red dashed line indicating that they were directly grown on $r$-sapphire with the full connectedness since there is no $h$-BN space layer. The garble-roof shape of both detached and undetached GaN domains is another indication of the thru-hole epitaxy verifying the crystallographic alignments of GaN domains with the underlying sapphire substrate.

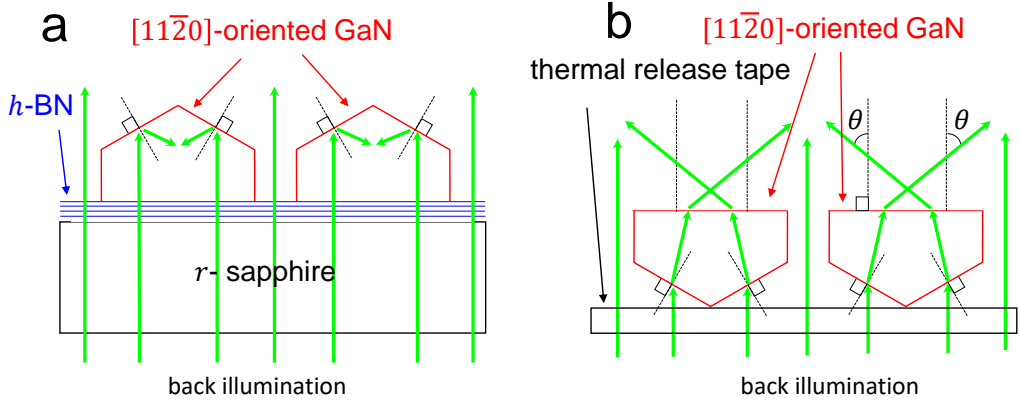

Supplementary Fig. 12: **Garble-roof shape: the reason why the [11$\bar{2}$0]-oriented GaN domains look dark.** a–b Ray-tracing diagrams for (a) as-grown and (b) detached GaN domains corresponding to the images shown in Supplementary Fig. 11a and c. The incident light in (a) is totally internally reflected with an incidence angle of 30°, which is larger than the critical angle of GaN in a full visible range. On the other hand, the incident light in (b) exits GaN after refraction twice with a final exit angle larger than the maximum acceptance angle in a full visible range determined by an objective lens. These are two reasons why [11$\bar{2}$0]-oriented GaN domains with the garble-roof shape in Supplementary Fig. 11 are dark and the other regions look bright in the back illumination configuration.

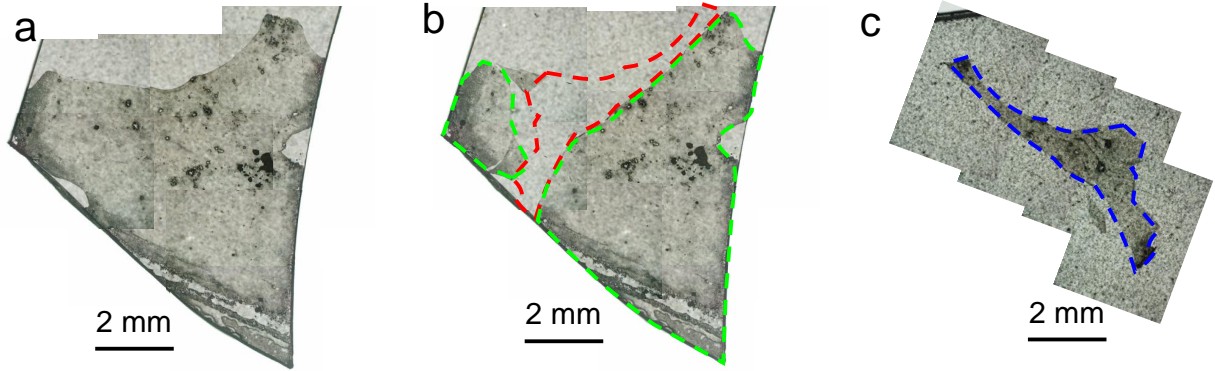

Supplementary Fig. 13: **Detachability near the critical connectedness.** **a**–**c** Stitched optical microscopy images of (**a**) as-grown GaN domains on a one-time transferred *h*-BN space layer on a *c*-sapphire substrate, (**b**) the *c*-sapphire substrate after the GaN domains were detached by using a handy thermal release tape, and (**c**) the detached GaN domains on the thermal release tape. The regions enclosed by the red and blue dashed lines represent the exposed *c*-sapphire and detached GaN. The region enclosed by the green dashed line represents the undetached GaN. In this case, the connectedness is below the critical value resulting in partial detachment of GaN. The extent of the connectedness can be easily reduced by increasing the number of transfers as shown in Supplementary Fig. 10, so that the detachability efficiency is much improved.

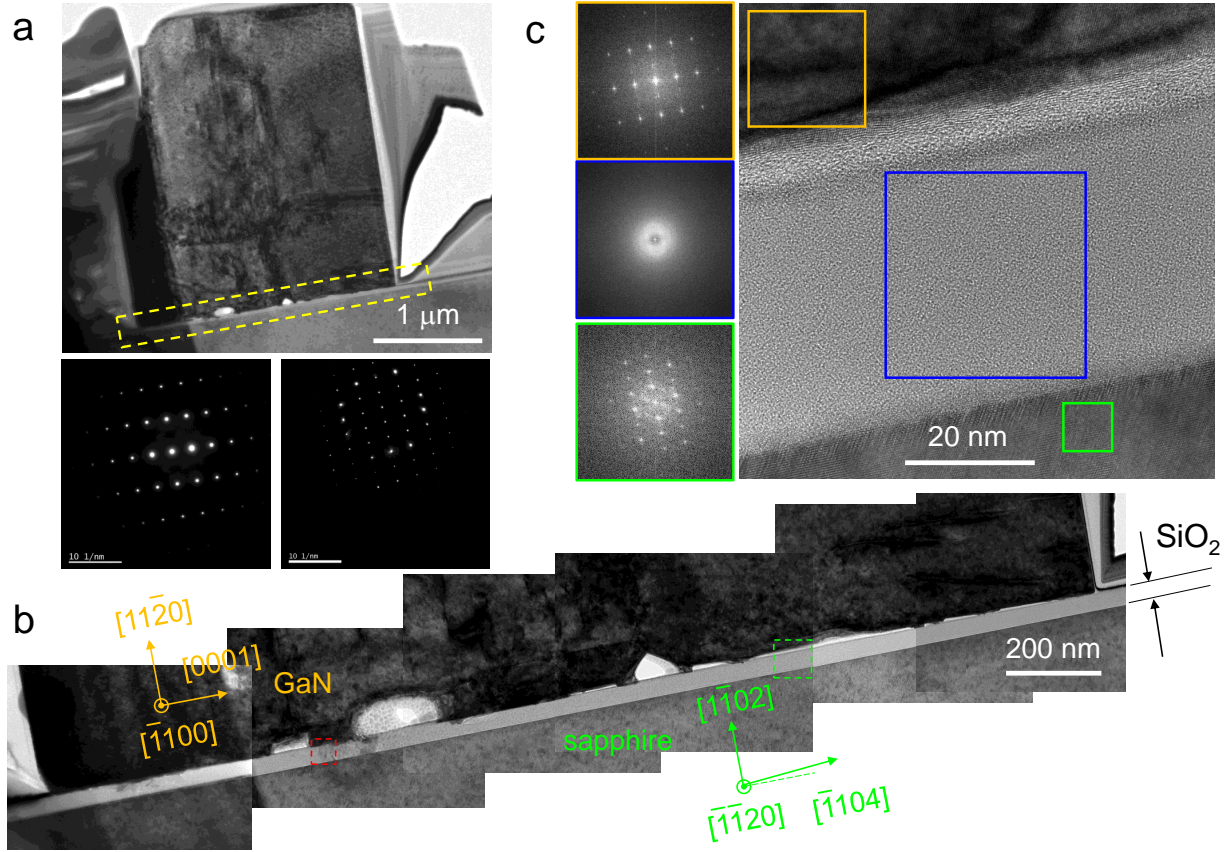

Supplementary Fig. 14: **TEM images across the interface of GaN/$h$-BN/SiO$_2$/$r$-sapphire.** **a** Cross-sectional TEM image of a single domain of GaN grown on $h$-BN/SiO$_2$/$r$-sapphire with selected area diffraction patterns of GaN (left) and sapphire (right). **b** Magnified stitched images of the interface region enclosed by the yellow dashed box shown in (**a**). **c** Magnified TEM image of the region enclosed by the green dashed box shown in (**b**) and the FFTs of the GaN, SiO$_2$, and the $r$-sapphire regions enclosed respectively by the orange, blue, and green boxes. It is clear that there is a $h$-BN space layer on top of the SiO$_2$ film. The high-resolution TEM image, taken from the region with a thru-hole enclosed by the red dashed box, is shown in Fig. 5d in the main text. The FFTs reveal that $[11\bar{2}0]$-oriented GaN is aligned with $r$-sapphire, suggesting the lateral overgrowth initiated by the thru-hole epitaxy. Note that the FFT of the SiO$_2$ region shows no crystallinity at all.

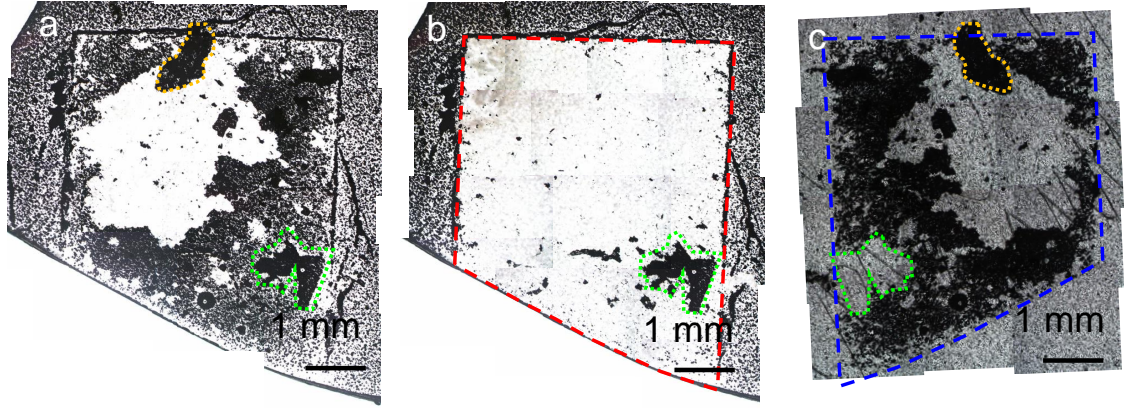

Supplementary Fig. 15: **TRT transfer of GaN grown on $h$-BN/SiO$_2$/$r$-sapphire.** **a–c** Stitched optical microscopy images of (**a**) as-grown GaN domains on a one-time transferred $h$-BN space layer on an SiO$_2$/$r$-sapphire substrate, (**b**) the $r$-sapphire substrate after the GaN domains were detached by using a handy thermal release tape, and (**c**) the detached GaN domains on the thermal release tape. As mentioned in Supplementary Fig. 11, the dark-looking GaN domains are all [11$\bar{2}$0]-oriented with the garble-roof shape. The white region in the middle of the sample in (**a**) is where GaN was self-separated during the cool-down process after the growth. The regions enclosed by the red and blue dashed lines represent the exposed $r$-sapphire and detached GaN. The regions enclosed by the green and yellow dotted lines represent the undetached and self-separated GaN domains. We speculate that this region enclosed by a green dashed line was not detached due to either the high connectedness or the direct bonding with SiO$_2$. Note that there are dark-looking GaN domains with a garble-roof shape outside the region enclosed by the red dashed line indicating that they were directly grown on SiO$_2$/$r$-sapphire without a $h$-BN space layer.



\end{document}